\documentclass[12pt,preprint]{aastex}
\usepackage{graphicx}
\usepackage{graphics}
\usepackage{amsmath}
\usepackage{enumerate}
\usepackage{longtable}
\usepackage{rotating}
\usepackage{natbib}

\begin{document}

\doublespace\normalfont

\title{Reaction Studies of Neutral Atomic $\mathbf{\rm C}$ with
  $\mathbf{\rm H_3^+}$ using a Merged Fast-Beams Apparatus}

\author{A.  P.  O'Connor\altaffilmark{1,2},
X.  Urbain\altaffilmark{3},
J.  St\"utzel\altaffilmark{1,4}, 
K.  A.  Miller\altaffilmark{1}, 
N.  de Ruette\altaffilmark{1}, 
M.  Garrido\altaffilmark{1}, \\
and D.  W.  Savin\altaffilmark{1}}
\altaffiltext{1}{Columbia Astrophysics Laboratory, Columbia
  University, New York, NY 10027, U. S. A. }
\altaffiltext{2}{Present address: Max-Planck Institute for Nuclear
  Physics, Heidelberg 69117, Germany}
\altaffiltext{3}{Institute of Condensed Matter and Nanosciences,
  Universit\'e Catholique de Louvain, Louvain-la-Neuve B-1348,
  Belgium}
\altaffiltext{4}{Present address: Bosch Engineering GmbH,
  Robert-Bosch-Allee 1, 74232 Abstatt, Germany}
\email{savin@astro.columbia.edu}

\begin{abstract}

We have investigated the chemistry of ${\rm C + H_3^+}$ forming
CH$^+$, CH$_2^+$, and CH$_3^+$.  These reactions are believed to be
some of the key gas-phase astrochemical processes initiating the
formation of organic molecules in molecular clouds.  For this work we
have constructed a novel merged fast-beams apparatus which overlaps a
beam of molecular ions onto a beam of ground-term neutral atoms.  Here
we describe the apparatus in detail and present cross section data for
forming CH$^+$ and CH$_2^+$ at relative energies from $\approx
9$~meV to $\approx20$ and 3~eV, respectively.  Measurements were
performed for statistically populated C$(^3P_J)$ in the ground term
reacting with hot H$_3^+$ (at an internal temperature of $\sim
2,550$~K).  Using these data we have derived rate coefficients for
translational temperatures from $\approx72$~K to $\approx2.3
\times 10^5$ and $3.4 \times 10^4$~K, respectively.  For the formation
of CH$_3^+$ we are able only to put an upper limit on the rate
coefficient.  Our results for CH$^+$ and CH$_2^+$ are in good
agreement with the mass-scaled results from a previous ion trap study
of ${\rm C + D_3^+}$, at a translational temperature of $\sim
1,000$~K.  That work also used statistically populated C$(^3P_J)$ but
internally cold D$_3^+$ ($\sim 77$~K).  The good agreement between the
two experiments implies that the internal excitation of the H$_3^+$ is
not significant so long as the reaction proceeds adiabatically.  At
300~K, the C fine-structure levels are predicted to be essentially
statistically populated, enabling us to compare our 
  translational temperature results to thermal equilibrium
calculations.  At this temperature our rate coefficient for forming
CH$^+$ lies a factor of $\sim 2.9$ below the Langevin rate coefficient
currently given in astrochemical databases and a factor of $\sim
1.8-3.3$ below the published  classical trajectory studies using
  quantum mechanical potential energy surfaces.  Our results for
CH$_2^+$ formation at 300~K are a factor of $\approx 26.7$, above
these semi-classical results.  Astrochemical databases do not
currently include this channel.  We also present a method for
converting our translational temperature results to thermal rate
coefficients for temperatures below $\approx 300$~K.  The results
indicated that CH$_2^+$ formation dominates over that of CH$^+$ at
temperatures $\lesssim 50$~K.

\end{abstract}

\keywords{Astrobiology -- Astrochemistry -- ISM: molecules -- 
Methods: laboratory -- Molecular data -- Molecular processes}

\section{Introduction}
\label{sec:Intro}

The first organic molecules are thought to have formed through
interstellar gas-phase chemistry when atomic carbon was ``fixed'' into
hydrocarbons.  Typical molecular cloud densities are so low that one
needs only consider binary collisions.  As a result, the initial
chemical network involved is rather simple, primarily consisting of
C$^+$ or C reacting with either H, H$_2$, or H$_3^+$
\citep{vanD98a,Herb08a}. The cosmic pathway from there to more complex
hydrocarbons and other organic molecules passes through the molecular
cations CH$_n^+$ ($n=1-3$). Understanding how C$^+$ and C react to
form these ions is therefore critical for modeling the origins of
organic chemistry.

The available pathways for the C$^+$ network have been well laid out
\citep{vanD98a,Herb08a}, even if significant uncertainties remain in
the actual rate coefficients \citep{Vasy08a,Wake09a,Wake10a}.  Naively
one would expect the hydrogen abstraction reaction
\begin{equation}
\label{eq:C+H2_HA}
{\rm C^+ + H_2 \to CH^+ + H}
\end{equation}
to be important. However, this process is endothermic by 0.4~eV and
does not go forward at the low temperatures typical of molecular
clouds. Instead, C$^+$ is thought primarily to undergo radiative
association via
\begin{equation}
\label{eq:C+H2_RA}
{\rm C^+ + H_2 \to CH_2^+ + photon}. 
\end{equation}
A discussion of theoretical and experimental work on this system
  can be found in \citet{Gerl92a} and \citet{Gerl08a} and references
  therein.  Hydrogen abstraction reactions of the product CH$_2^+$
with H$_2$ can then form CH$_3^+$.   Recent experimental
studies of hydrogen abstraction involving CH$^+$ and CH$_2^+$ have
been published by \citet{Gerl11a}.

The corresponding C chemistry is expected to be dominated by reactions
with H$_3^+$, but the network is much more uncertain.  Astrochemical
databases currently include the proton transfer process
\citep{KIDA12,UMIST13}
\begin{equation}
\label{eq:CH3+_pT}
{\rm C + H_3^+ \to CH^+ + H_2}. 
\end{equation}
Classical trajectory studies using quantum mechanical potential
  energy surfaces (PESs) have been used to calculate the thermal rate
coefficient for this reaction in the range of $10-300$~K
\citep{Talb91a,Bett98a,Bett01a}.  That semi-classical work takes
into account the temperature dependence of both the 
  translational motion and the internal energy of the C atom but
assumes the H$_3^+$ ion is in the ground state.

Ion trap measurements for
\begin{equation}
\label{eq:CD3+_pnT}
{\rm C + D_3^+ \to CD^+ + D_2},
\end{equation}
have been performed by \citet{Savi05a} at an estimated  translational temperature of $\sim 1,000$~K (see
Section~\ref{sec.kineticrate} for a discussion of both the
  translational and internal temperatures in their work).  Throughout
  this paper, we use the term ``translational temperature'' to refer
  to the reaction center-of-mass velocity distribution when it is
  described by a Maxwell-Boltzmann distribution.  The internal
  energies of the reactants, however, are not necessarily in thermal
  equilibrium.  When both the translational temperature and internal
  energies are in thermal equilibrium, we use the term ``thermal''.

Comparing theory and experiment for these analogous reactions touches
on a major issue in astrochemistry, namely how to convert data
between isotopologues of a collision system.  Two approaches are
commonly used. Some researchers assume the rate coefficients are the
same, independent of the isotopologues involved
\cite[e.g.,][]{Rodg96a,Alber13a}.  Others use the Langevin theory
\citep{Giou58a} and scale by the square root of the ratio of the
reduced masses \citep[e.g.,][]{Stan98a,Gay11a}.  These two approaches
result in a multiplicative scaling factor of 1 and 1.29, respectively,
for the \citet{Savi05a} data. However, even taking this into account,
theory and experiment have still not converged in either magnitude or
temperature dependence.  The theoretical calculations differ from one
another by a factor of about 1.7. Including the error bars on the
laboratory work, the published rate coefficients for this reaction
span nearly an order of magnitude; though it is unclear if this
represents a temperature dependence in the reaction or is a true
discrepancy. But even assuming only a factor of 2 uncertainty,
astrochemical sensitivity studies still find that improving the
accuracy of this rate coefficient is of critical importance for
reliably matching model predictions to observations
\citep{Vasy08a,Wake09a,Wake10a}.

Current astrochemical databases, however, do not consider the
possibility of the additional ${\rm C + H_3^+}$ reaction channels
\begin{equation}
\label{eq:CH3+_ppT}
{\rm C + H_3^+} \to {\rm CH_2^+ + H}
\end{equation}
and
\begin{equation}
{\rm C + H_3^+} \to {\rm CH_3^+ + photon}\label{eq:CH3+_RA}.
\end{equation}
If these reactions are fast enough, then they could result in an
increased efficiency for the gas-phase formation of complex
hydrocarbons. However, the theoretical predictions and experimental
findings for these two reactions are not in agreement.

For reaction~(\ref{eq:CH3+_ppT}), \citet{Talb91a} calculate that it
possesses a significant activation energy and will not proceed at
typical molecular cloud temperatures. \citet{Bett98a,Bett01a}, though,
find that the reaction proceeds with no barrier at a rate a factor of
$\sim 60-110$ smaller than that for reaction~(\ref{eq:CH3+_pT}). That
this channel is open is supported by the experimental work of
\citet{Savi05a} on the analogous system
\begin{equation}
\label{eq:CD3+_pnpnT}
{\rm C + D_3^+} \to {\rm CD_2^+ + D}. 
\end{equation}
However, they measure a rate coefficient that is only a factor of
$\sim 2$ smaller than that for reaction~(\ref{eq:CD3+_pnT}).

As for reaction~(\ref{eq:CH3+_RA}), it was not considered by either
\citet{Talb91a} or \citet{Bett98a,Bett01a}. Quite likely that is
because in binary collisions the process can only proceed via
radiative association.  Such reactions typically have rate
coefficients many orders of magnitude smaller than processes such as
reactions~(\ref{eq:CH3+_pT}) and (\ref{eq:CD3+_pnT}), which are
expected to proceed with near Langevin rate coefficients
\citep[e.g.,][]{Herb08a}. Surprisingly though, \citet{Savi05a}
measured a rate coefficient for the analogous process
\begin{equation}
\label{eq:CD3+_RA}
{\rm C + D_3^+} \to {\rm CD_3^+}. 
\end{equation}
and found that it is a factor of only $\sim 12$ smaller than that for
reaction~(\ref{eq:CD3+_pnT}).

It is clear that additional research is needed to improve our
understanding of the ${\rm C + H_3^+}$ reaction complex. However, to
accomplish this goal, there are formidable challenges both
theoretically and experimentally.

Astrochemical databases use the Langevin value for the ${\rm C +
  H_3^+}$ reaction system.  The only detailed calculations of which we
are aware for this system are the semi-classical results of
\citet{Talb91a} and \citet{Bett98a,Bett01a}. Fully quantum mechanical
scattering calculations for ion-neutral collision systems with 4 or
more atoms appear to be just beyond current theoretical
capabilities. The deep potential wells require large reactant and
product basis sets. Accurate long-range potentials are needed as they
are predicted to drive the reaction process.  Additionally, multiple
electronic surfaces may be involved along with non-adiabatic coupling
between the surfaces. Brief reviews of the field can be found in
\citet{Alth03a} and \citet{Bowm11a}. In the meanwhile, the
state-of-the-art seems to be represented by the work of
\citet{Klip10a}, which is a combination of transition state theory,
classical trajectory simulations, and master equation analysis. They
calculated the ${\rm O}(^3P) + {\rm H_3^+}$ and ${\rm CO + H_3^+}$
systems; but we are unaware of any similar work on ${\rm C}(^3P) +
{\rm H_3^+}$.

Experimentally, studies of cross sections and rate coefficients for
reactions of C with molecular ions are extremely difficult. Part of
the difficulty has to do with the challenge of generating beams of
neutral atomic C. Standard experimental techniques for measuring
ion-neutral reactions, such as flowing afterglows and related
approaches, cannot generate sufficient amounts of neutral atomic C due
to its high reactivity (A. Viaggiano, private communication). Laser
ablation can produce beams of atomic C \citep{Kais95a,Gu06a},
  but have yet to be used in an experimental configuration that can
  generate cross sections or rate coefficients \citep{Wils12a}.  In
fact, we are aware of only two published laboratory measurements of
cross sections or rate coefficients for reactions of C with molecular
ions \citep{Schu83a,Savi05a}.

The approach of \citet{Schu83a} was to send a fast beam of C$^+$
through a gas cell, neutralizing a portion of the beam through
electron capture. After the cell, any remaining C$^+$ was magnetically
removed, leaving a beam of neutral C. Merging an internally hot
D$_2^+$ beam with the neutral beam, they studied the reaction
\begin{equation}
\label{eq:CD2+_DT}
{\rm C + D_2^+ \to CD^+ + D}. 
\end{equation}
Because both beams were fast, standard laboratory methods could be
used to characterize the parent beam profiles and particle currents
and also to detect the product ions. Hence it was possible to perform
absolute cross section measurements. However, the neutral C beam
contained an unknown mixture of ground state and metastable levels of
C with internal energies of up to $\sim 4.2$~eV. This limits the
ability to make an unambiguous comparison of the results with
theoretical calculations. Moreover, it also prevents the use of the
results for astrochemistry where any neutral C atoms are expected to
be in the ground term.

Another experimental approach taken is that of \citet{Savi05a}. They
used heated graphite rods to create an effusive beam consisting of a
mixture of C, C$_2$, and C$_3$, which flowed into an ion trap
containing internally cold D$_3^+$. With their apparatus they
investigated reactions~(\ref{eq:CD3+_pnT}), (\ref{eq:CD3+_pnpnT}), and
(\ref{eq:CD3+_RA}). Rate coefficients were determined by measuring the
trapped parent and daughter ion populations versus time. However,
there are a number of drawbacks to this method: (a) the beam is not
pure and the C$_n$ impurities ($n \ge 2$) can react with the trapped
ions, potentially affecting the results; (b) the carbon source emits
vacuum ultraviolet radiation which can cause ionization in the trap
and alter the chemistry occurring; (c) the carbon beam is emitted in
bursts and has an unknown density which varies spatially and
temporally, complicating the determination of the neutral-ion overlap
and the extracted rate coefficient; (d) the energy of the carbon atoms
and the energy spread of the effusive beam are both highly uncertain;
and (e) the $\sim 1,000$~K translational temperature of the
experiment is a factor of $\gtrsim 100$ higher than typical molecular
cloud temperatures.  However, perhaps the biggest issue is that (f)
trapped ionic end products can undergo subsequent parasitic reactions
with either parent-beam or background-gas neutrals, complicating the
interpretation and analysis of the data. This last point is the reason
that only lower limits were given for the error bars on their
measurements of reactions~(\ref{eq:CD3+_pnpnT}) and
(\ref{eq:CD3+_RA}).

It is clear that there is a need for an improved ability to study
reactions of neutral atomic C with molecular ions. Here we describe a
novel, merged fast-beams apparatus that we have developed to study
such reactions. For our proof-of-principle studies we have
investigated reactions~(\ref{eq:CH3+_pT}), (\ref{eq:CH3+_ppT}), and
(\ref{eq:CH3+_RA}). As we describe below, our approach overcomes many
of the limitations of \citet{Schu83a} and \citet{Savi05a}.

The rest of this paper is organized as follows. In
Section~\ref{sec:Experimental} we review our experimental approach. A
detailed description of the apparatus is provided in
Section~\ref{sec:Apparatus}. Section~\ref{sec:energies} discusses the
beam profile measurements and simulations used to determine the
relative energies.  Section~\ref{sec:Signal} outlines how the signal
was determined while Section~\ref{sec:DAQ} reviews the data
acquisition method. Uncertainties and data averaging are briefly
discussed in Section~\ref{sec:Uncert}. Results are presented in
Section~\ref{sec:results} and discussed in
Section~\ref{sec:discussion}.  Some astrochemical implications are
explored in Section~\ref{sec:astro}.  In Section~\ref{sec:summary} we
summarize our findings.

\section{Experimental Approach}
\label{sec:Experimental}

\begin{figure}[!p]
\centering
\includegraphics[width=1\textwidth]{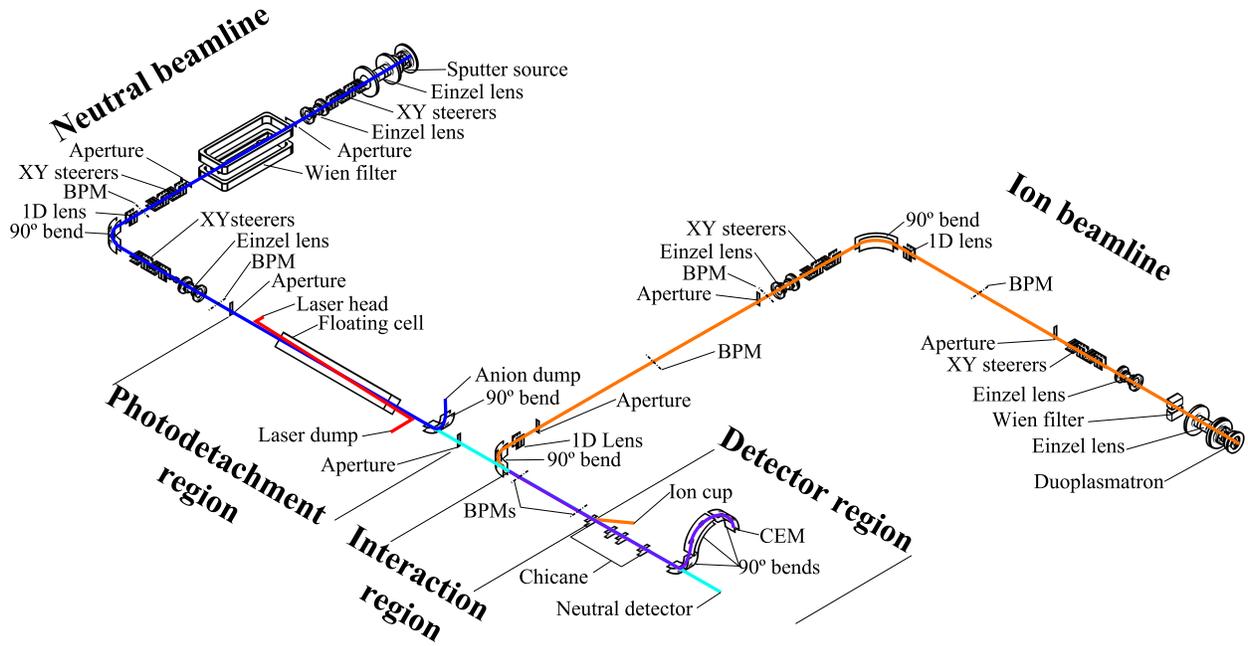} 
\caption{Overview of the merged-beams apparatus.}
\label{Fig:apparatus}
\end{figure}

We have developed a versatile merged fast-beams apparatus
capable of studying a range of chemical processes involving neutral
atoms or molecules reacting with atomic or molecular ions (see
Figure~\ref{Fig:apparatus}). As currently configured, the system is
designed to detect charged daughter products which are heavier than
either of the parent beams, such as reactions~(\ref{eq:CH3+_pT}),
(\ref{eq:CH3+_ppT}), and (\ref{eq:CH3+_RA}).

The neutral beam portion of the apparatus begins with a Cs-ion sputter
source to generate singly charged anions. By floating the source
cathode to a negative potential of $U_{\rm s}$, we non-selectively
extract anions and generate a fast beam with a laboratory kinetic
  energy of $-eU_{\rm s}$, where $e$ is the unit charge. A Wien
filter is used to purify the beam by selecting for the charge-to-mass
ratio of the desired ion, which is then directed into an electrically
isolated cell, floated to a potential of $U_{\rm f}$.  Inside this
floating cell the anions, now with an energy of $-e(U_{\rm s} - U_{\rm
  f})$, are crossed by a laser beam which photodetaches a fraction of
the beam. After exiting the photodetachment chamber, the anions are
electrostatically deflected into a Faraday cup, resulting in a neutral
beam with a kinetic energy of $E_{\rm n} = -e(U_{\rm s} - U_{\rm
  f})$.  For the second beam in the experiment, we use a duoplasmatron
ion source floated to a positive potential of $U_{\rm d}$. We select
the desired cations with a Wien filter. The resulting ion beam  kinetic energy is $E_{\rm i} = eU_{\rm d}$. An electrostatic
deflector is used to merge the ions with the neutral beam.

For mono-energetic beams, the relative energy in the
  center-of-mass is given by
\citep{Brou83a}
\begin{equation}
E_\mathrm{r} = 
\mu 
\left(
\frac{E_\mathrm{n}}{m_\mathrm{n}} + \frac{E_\mathrm{i}} {m_\mathrm{i}}
- 2\sqrt{\frac{E_\mathrm{n}E_\mathrm{i}}{m_\mathrm{n}m_\mathrm{i}}}\cos\theta
\right).
\label{Eqn.brou}
\end{equation}
where $m_{\rm n}$ and $m_{\rm i}$ are the neutral and ion masses,
respectively; $\mu = m_{\rm n}m_{\rm i}/(m_{\rm n} + m_{\rm i})$ is
the reduced mass of the collision system; and $\theta$ is the
intersection angle. This relative energy between the two beams
can be controlled by varying either the source potentials or the
floating cell potential or some combination thereof.

The beginning of the interaction region is determined by the point
inside the electrostatic deflector where the ions merge with the
neutral beam. A beam profile monitor (BPM) is mounted near the
beginning of the interaction region and another near the end. These
are used to measure the horizontal and vertical profile of each beam
and determine their overlap. During data acquisition, both beams are
chopped on and off, out of phase. This enables us to extract the
desired signal from various backgrounds.  Parent beam densities
  of $\sim 10^4-10^5$~cm$^{-3}$ help reduce to insignificant any
  possible effects of three-body and parasitic reactions.

\begin{figure}[!p]
\centering
\includegraphics[width=1\textwidth]{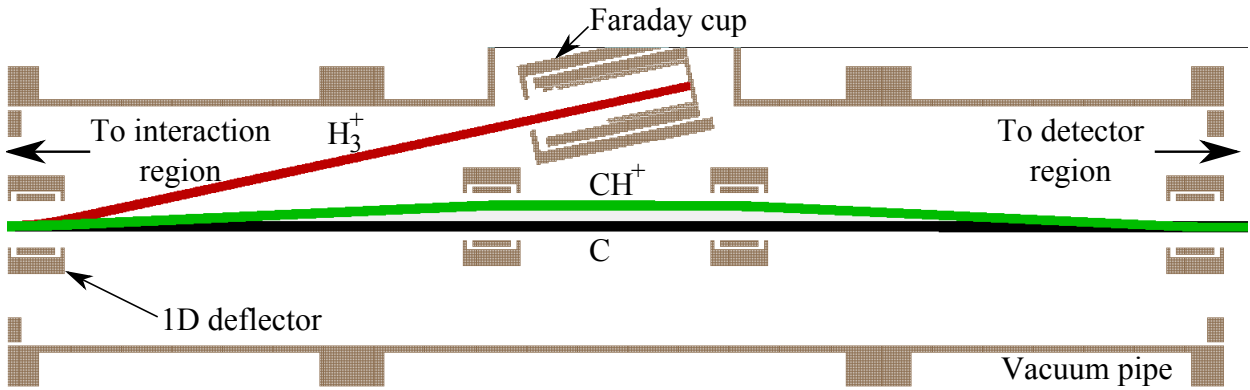} 
\caption{SIMION model of the chicane showing the demerging of the ion
  beams using a series of parallel-plate 1D deflectors, only the first
  one of which is labeled. The H$_3^+$ is directed upwards into a
  Faraday cup and the current recorded. The CH$^+$ beam is then
  re-merged with the C beam and directed into an electrostatic energy
  analyzer.  Any CH$_2^+$ or CH$_3^+$ will be deflected less strongly
  than the CH$^+$ beam.}
\label{Fig:chicane}
\end{figure}

The end of the interaction region is determined by an electrostatic
chicane (Figure~\ref{Fig:chicane}) which sends the parent ions into a
Faraday cup, while the parent neutral beam continues ballistically and
any heavier, charged daughter products are directed into an
electrostatic energy analyzer. A hole in this analyzer allows the
neutral beam to pass through into a detector which measures the
neutral beam particle current. The desired product ion is selected
based on its final kinetic energy and directed into a channel
electron multiplier (CEM) operated in pulse counting mode. We note
that, as a result of the high beam velocities in the laboratory frame,
the angular spread of the reaction products is strongly compressed in
the forward direction onto a small surface area. This enables us to
use standard detection techniques to collect the signal ions emitted
over the entire 4$\pi$ steradians in the center-of-mass frame.

Experimentally we measure the reaction cross section $\sigma$ times
the relative velocity $v_{\rm r}$ convolved with the relative
  velocity spread of the experiment.  This merged-beams rate
coefficient can be written as \citep[e.g.,][]{Bruh10b}
\begin{equation}
\left<\sigma v_r \right> = 
\frac{S}{T_{\rm a}T_{\rm g} \eta}
\frac{e^2v_{\rm n}v_{\rm i}}{I_{\rm n}I_{\rm i}}
\frac{1}{L\left<\Omega(z)\right>}. 
\label{eq:rate}
\end{equation}
Here $S$ is the count rate measured at the CEM, $T_{\rm a}$ is the
transmittance of the analyzer for the selected daughter product,
$T_{\rm g}$ is the geometric transmittance of the grid in front of the
CEM, $\eta$ is the CEM efficiency, $v_{\rm n}$ is the neutral beam
velocity, $v_{\rm i}$ is that of the ion beam, $I_{\rm n}$ is the
neutral particle current measured in amperes, $I_{\rm i}$ is the ion
current, and $L$ is the length of the interaction region. The term
$\left<\Omega(z)\right>$ is the average overlap integral in the
interaction region and is discussed further in Section~\ref{sec:Over}.

Because we measure all of the quantities on the right hand side of
Equation~(\ref{eq:rate}), we are able to report absolute measurements.
From a combination of beam profile measurements and trajectory models,
we are able to determine the interaction relative energy
spread. This enables us to deconvolve our results to generate cross
sections which can then be re-convolved with a Maxwell-Boltzmann
distribution to generate a translational temperature rate
coefficient.

\section{Apparatus Description}
\label{sec:Apparatus}

The apparatus design is based, in part, on that of our self-merged
fast-beams apparatus which has already been described in a
series of papers \citep{Krec10c,Bruh10a,Bruh10b}. For example, many of
the ion optics for steering and focusing are similar to those used
previously. Here we describe only those additional details specific to
this new apparatus.

\subsection{Carbon Beam Line}
\label{sec:carbon}

We generate a beam of C$^-$ using a Cs-ion sputter source in
combination with a Wien filter to remove all other unwanted negative
particles. The Wien filter can readily resolve $^{12}$C$^-$ from both
$^{13}$C$^-$ and $^{12}$C$^1$H$^{-}$, thereby enabling us to generate
an isotopically pure $^{12}$C$^-$ beam. Generally we operate the
  beam at a relative energy of $E_{\rm C^-} \approx 28$~keV. The
full width half maximum (FWHM) energy spread for a C$^-$ beam from a
sputter source is typically $\sim 15$~eV \citep{Douc77a,Douc77b}.
Operating pressures in the region of the ion source are $\sim
10^{-6}$~Torr.

C$^-$ is one of the rare atomic anions which possess more than one
stable bound level lying below the first detachment threshold: a
ground $2s^22p^3\ ^4S_{3/2}$ level and two excited $2s^22p^3 \ ^2D_J$
levels ($J = 5/2$ and 3/2). However, sputter sources have been shown
to produce insignificant populations of the $^2D_J$ excited levels
\citep{Sche98a,Taka07a}.  Thus we produce an essentially pure beam of
ground level C$^-$.

After exiting the Wien filter, the beam is directed into a 90$^\circ$
electrostatic cylindrical deflector. This deflection prevents any
neutral particles or ultraviolet photons emitted by the source from
having a direct path into the interaction region. Typical pressures in
the deflector are $\sim 10^{-8}$~Torr. Nominal C$^-$ currents after
this deflector are $\sim 1.8$~$\mu$A.

The anion beam is then directed through a 5~mm diameter circular
aperture and continues essentially ballistically. Along this second
leg of the carbon beamline, we have installed horizontal and vertical
solenoid coils with rectangular cross sections. These enable us to
largely cancel out the effects of the Earth's magnetic field, which
would generate an unwanted deflection of the anion beam.

After the 5~mm aperture, the anions enter into the floating cell,
housed in what we call the photodetachment chamber. Outside of this
chamber we use a diode laser to generate an 808-nm (1.53-eV) laser
beam with $\approx1.8$~kW of power. Using lenses and mirrors external
to the vacuum chamber, the beam is directed into the floating cell.
Near the center of the overlap region with the anion beam, the laser
light is brought to an oval-shaped focus where 90\% of the power lies
in an area of 9.4~mm in the horizontal direction and 11.6~mm in the
vertical direction.  The laser crosses the anions at an angle of $\phi
\approx 2.74^\circ$. The laser beam exits the chamber a distance of
2008~mm from the entrance and is directed into a water-cooled power
meter, which we monitor during data acquisition.

Based on the known photodetachment cross section
\citep{Sema62a,Zhou04a}, laser power, and anion velocity, and using
the expected beam shapes and overlap geometry, we estimate that $\sim
4\%$ of the C$^-$ beam is photodetached into ground term C($^3P)$,
though not all of the beam passes through subsequent apertures in the
system. The photon energy and flux are insufficient to photodetach
into higher lying levels of atomic C.

\begin{figure}[!p]
\centering
\includegraphics[width=1\textwidth]{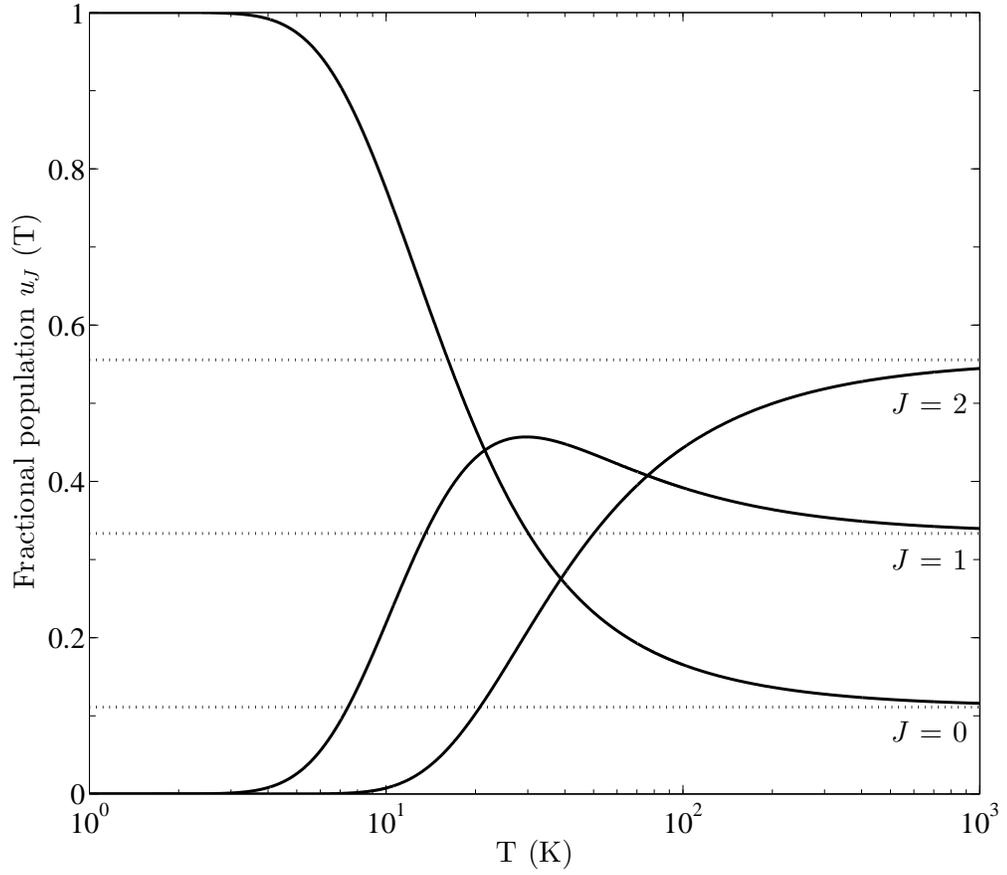}
\caption{Fractional population of the $^3$P$_J$ levels for a
  thermal distribution (solid curves) and a statistical distribution
  (dotted lines).}
\label{Fig:Clevels}
\end{figure}

Based on previous photodetachment studies, we expect to statistically
populate all three fine-structure $J$ levels of the $^3P$ ground term
\citep{Sche98a}. The $J = 1$ and 2 levels lie above the $J = 0$ ground
level by energies of $E_J = 2.0$ and 5.4~meV, respectively.  We have
calculated the thermal population of the $J$ levels using the
partition functions
\begin{equation}
\label{eq:partition}
u_J = \frac{g_J e^{(-E_J/k_{\rm B}T)}}{\sum_J g_J
    e^{(-E_J/k_{\rm B}T)}},
\end{equation}
where $g_J = 2J + 1$ is the statistical weight of level $J$.  The
results are shown in Figure~\ref{Fig:Clevels}, which indicates that
the fine-structure population reaches a statistical distribution for
temperatures $\gtrsim 300$~K.  We will return to this issue in
Section~\ref{sec:discussion}.

The neutral beam exits the floating cell with a relative
energy of $E_{\rm C} = E_{\rm C^-} + e U_{\rm f}$. Both the anion and
neutral beams then enter a $90^\circ$ cylindrical deflector. The
remaining C$^-$ ions are electrostatically deflected into a Faraday
cup which collects the anion beam. Typical C$^-$ currents measured at
this point are on the order of 1~$\mu$A.

The neutral beam then passes through a 12~mm diameter aperture in the
outer plate of this cylindrical deflector and after that through a
second 5 mm aperture, a distance of 3168~mm downstream from the first.
The separation of these two 5~mm apertures geometrically limits the
divergence of the C beam to a maximum half angle of 1.57~mrad. The
beam continues into yet another cylindrical electrostatic deflector
and passes through a 12~mm diameter hole in the outer plate of that
deflector. This deflector is used to merge the molecular ions with the
neutral beam. For data acquisition, we chop the C beam on and off by
chopping the laser beam. The switching time of the laser is on the
order of a few hundred~ns.

\subsection{Molecular Beam Line}
\label{sec:molecular}

The molecular ions are formed using a duoplasmatron. We extract a beam
of cations from the source and use a Wien filter to select the desired
H$_3^+$ ions. The typical beam kinetic energy is $E_{\rm
  H_3^+} \approx 7.05$~keV, chosen to match the laboratory
velocity of the $\approx 28$~keV C beam. The typical energy
spread from a duoplasmatron has a FWHM of $\sim 10$~eV
\citep{Aber67a}. The pressure in the vicinity of the source is $\sim
10^{-6}$~Torr.

The vibrational and rotational temperatures of $\mathrm{H_3^+}$ formed
in a duoplasmatron may be quite substantial. This is due to the
formation mechanism, namely, proton transfer between H$^+_2$ and
H$_2$, at least one of which is typically vibrationally excited in the
discharge. Internal energies of $E \sim 1$~eV are inferred by
\citet{Anic84a}.  These are supported by experimental studies which
found internal energies ranging between $\sim 0.5-1$~eV, with a
generally decreasing internal energy as the source load pressure
increases (X.\ Urbain, private communication).  We convert this to an
internal temperature using the theoretical partition function results
of \citet{Kyla11a}, specifically their Equation~(8).  Based on this,
we estimate that our H$_3^+$ internal temperature lies between $\sim
2,200$ and 3,400~K, which is below the 4,000~K dissociation
temperature of the system.  The internal temperature is higher than
that expected for many astrochemical environments.  However, as we
show later, this internal excitation appears not to have a significant
effect on our measurements down to temperatures of at least $\approx
300$~K.  Still, in future work we hope to explore the possibility of
installing a cold molecular ion source on our system.

As in the C beamline, the H$_3^+$ beam is deflected 90$^\circ$ into a
second leg which is surrounded by rectangular shaped solenoid coils to
cancel the magnetic field of the Earth in the horizontal and vertical
directions. The beam is then directed into a drift region defined by
two 5~mm apertures separated by a distance of 3069~mm. The drift
region contains only two electrostatic ion optics, both just prior to
the second aperture.  The first is a horizontal electrode, dubbed the
``kicker''.  This is used to adjust the vertical angle of the beam
entering the 90$^\circ$ deflector which merges them onto the neutrals.
With this we are able to reduce the vertical angle between the ion
beam and the neutral beam in the interaction region.  The second
electrode is a horizontal plate opposite the kicker which we use to
chop the beam, allowing or preventing the beam from being sent into
the merger.  The potential on this electrode is controlled using a
fast high voltage switch with a switching time of better than 100
ns. This enables us to chop the H$_3^+$ beam on and off during data
acquisition.

Typical H$_3^+$ currents exiting the drift region are $\sim 250$~nA.
The divergence of the H$_3^+$ beam at this point is geometrically
constrained by the collimating apertures, which limit the maximum half
angle to 1.62~mrad. The beam then passes through a one-dimensional
(1D) electrostatic lens and into a 90$^\circ$ electrostatic
cylindrical deflector which merges the cations with the neutrals. The
1D lens is used to compensate for the focusing effects of the
cylindrical deflector in the horizontal or merging plane. The
divergence of the beam grows somewhat due to self-repulsion from space
charge effects within the beam and due to focusing effects from the
merging cylindrical deflector. Typical pressures in the beam merger
are $\sim 10^{-8}$~Torr.

\subsection{Interaction Region}
\label{sec:interaction}

The H$_3^+$ beam is brought horizontally onto the neutral C beam, near
the exit of the beam merger. The two beams then co-propagate for some
distance with a low relative velocity. Within this interaction region,
some of the parent cations and neutrals undergo chemical reactions,
generating daughter molecular ions.

The kinetic energy of the daughter ions is essentially the
sum of $E_{\rm C}$ plus the product of the H$^+_3$ kinetic
energy per amu ($\sim 2.35$~keV amu$^{-1}$) times the mass in amu,
transferred from the H$_3^+$.  For matched parent-beam 
  laboratory velocities, this typically corresponds to $\approx
30.35$~keV for forming CH$^+$, $\approx 32.70$~keV for CH$_2^+$, and
$\approx 35.05$~keV for CH$_3^+$.  These are the end products for
reactions~(\ref{eq:CH3+_pT}), (\ref{eq:CH3+_ppT}), and
(\ref{eq:CH3+_RA}), respectively. The dominant background ion is the
$\sim 28$~keV C$^+$ formed when the parent C beam is ionized by
collisions with residual gas in the vacuum system.

The parent beams are demerged using an electrostatic
chicane. Trajectory studies using the ion optics package
SIMION\footnote{www.simion.com} indicate that the overlap length of
the two beams is $1215 \pm 25$~mm. This distance includes both the
merging and demerging regions, which make up less than 6\% of the
interaction length. The beams merge with an initial angle in the
horizontal plane of $0.21 \pm 0.01$~rad and are brought parallel
within a distance 30~mm. The demerging occurs in the vertical
direction over a distance of 38~mm with a final angle between the
beams of $0.19 \pm 0.01$~rad. Here and throughout all uncertainties
are quoted at an estimated $1\sigma$ statistical confidence level.

\begin{figure}[!p]
\centering
\includegraphics[width=1\textwidth]{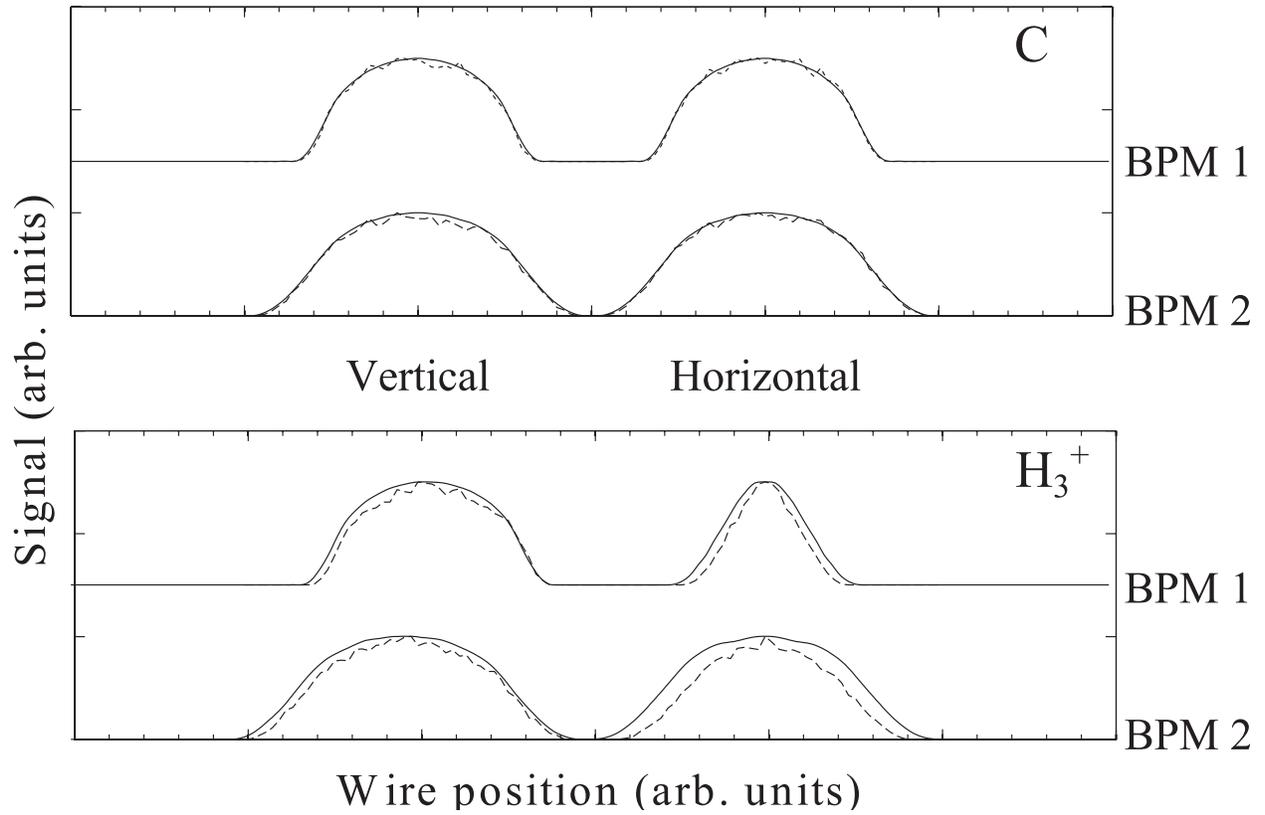} 
\caption{Comparison between the experimental profiles (dashed lines)
  and simulated profiles (solid lines) at BPM 1 and 2, respectively
  positioned 280 and 1090 mm from the start of the interaction
  region. Profiles are shown for both the C and H$_3^+$ beams. }
\label{Fig:Profiles}
\end{figure}

The profiles of the C and H$_3^+$ beams are measured at distances of
$280$ and $1090$~mm from the beginning of the interaction region
(Figure~\ref{Fig:Profiles}).  A retractable Faraday cup near the
middle of the interaction region can be used to measure ion beam
currents. The operating pressure in the interaction region is
generally $\sim 10^{-8}$~Torr. To minimize any deflection of the
parent and daughter product beams due to external magnetic fields, the
interaction region is shielded using a series of solenoid coils,
similar to the configuration in both the C and H$_3^+$ legs.

\subsection{Signal Detection}
\label{sec:detector}

One of the challenges of this research is separating the daughter ions
from the parent beams as well as from any unwanted background. We
accomplish the desired discrimination using electrostatic ion optics,
which allows us to analyze the beams based on their kinetic
energy.

The parent H$_3^+$ beam is demerged from the C beam by the first pair
of deflector plates in the chicane and directed with essentially 100\%
efficiency into a Faraday cup where the current is measured during
data acquisition. Typical currents are $\sim 200$~nA. This
  corresponds to H$_3^+$ beam densities of $\sim 10^5$~cm$^{-3}$.
The subsequent three pairs of deflector plates in the chicane bring
the product ions and background C$^+$ again onto the path of the
neutral C beam.

The remaining neutrals and ions then continue into an electrostatic
energy analyzer which consists of three 90$^\circ$ cylindrical
deflectors in series. The neutral beam passes unaffected through a
12~mm diameter hole in the outer plate of the lower cylindrical
deflector (LCD), and continues into a neutral detector which we
monitor during data acquisition. The calibration of this detector is
described in Section~\ref{sec:neutralcurrent}.

The cations are deflected vertically by the LCD, which directs the
desired ions into the middle cylindrical deflector (MCD). The MCD
deflects these ions perpendicular to their trajectories before and
after the LCD. There is a hole in the outer plate of the MCD, behind
which a Faraday cup is mounted. Thus, with no voltage on the MCD, we
can measure the beam current at this point. Dubbed the upper cup, the
transmittance from the interaction region to this Faraday cup is
$T_{\rm u} = 0.80 \pm 0.02$.

The upper cylindrical deflector (UCD) bends the selected ions another
90$^\circ$ in the same plane as that of the MCD, for a total of
180$^\circ$ in a single plane. We found that this 180$^\circ$
deflection was necessary in order to electrostatically separate C$^+$,
CH$^+$, CH$_2^+$, and CH$_3^+$ from each other. At the exit of the
UCD, the selected ions are moving downwards in the laboratory.

Ions which are discriminated against will strike somewhere on the
inside of the LCD, MCD, or UCD. The deflector plates of all three are
coated with a fine layer of graphite to minimize both photon and
secondary charged particle emission resulting from these undesired
ions hitting the surfaces. Typical operating pressures, as measured
near the MCD, are on the order of $\sim 10^{-9}$~Torr.

The transmittance through the Chicane, LCD, MCD, and UCD was optimized
using a proxy C$^-$ beam at kinetic energies similar to
those predicted for the product CH$_n^+$ ions and with inverted plate
voltage polarities.  This was carried out prior to data acquisition by
mounting a Faraday cup at the exit of the UCD. The transmittance was
found to be $T_{\rm a} = 0.73 \pm 0.02$.

For data collection we installed a CEM at the exit of the UCD. Ions
were detected in single particle counting mode. In front of the CEM we
have mounted a grid with a geometric transmittance of $T_{\rm g} =
0.90 \pm 0.01$. A voltage of $-200$~V is applied to the grid to repel
negative particles produced in any of the cylindrical deflectors.  The
CEM particle detection efficiency is discussed below in
Section~\ref{sec:cemefficiency}.

\subsection{Neutral Current Measurement}
\label{sec:neutralcurrent}

Typical neutral particle currents, as measured in amperes, are $\sim
30$~nA.  This corresponds to atomic C beam densities of $\sim
  1.4 \times 10^4$~cm$^{-3}$.  Beam fluctuations during data acquisition
introduce an $\approx 5\%$ uncertainty in the measured $I_{\rm C}$.
These are treated as a statistical uncertainty.

Neutral currents are measured on a particle collecting cup which can
be externally configured either as a Faraday cup to measure ion
currents directly or as a neutral detector to measure neutral particle
currents via secondary negative particle emission. The transmission
efficiency of a neutral beam from the interaction region to this cup
was measured using a 28~keV C$^{-}$ proxy beam and found to be $T_{\rm
  n} = 0.94 \pm 0.02$.

The neutral C particle current, as measured in amperes, is given by
\begin{equation}
I_{\rm C} = \frac{I_{\rm ND}}{\gamma T_\mathrm{n}}. 
\label{eq:C}
\end{equation}
Here $I_{\rm ND}$ is the current measured on the neutral detector and
$\gamma$ is the secondary negative particle emission coefficient for
$\sim 28$~keV C striking the detector.

We used collisional detachment of C$^-$ to determine the $\gamma$ of
the neutral detector. Helium was leaked into the chicane. As C$^-$
passed through the He, single and double electron detachment formed C
and C$^+$, respectively. Triple electron detachment was found to form
insignificant currents of C$^{++}$ for the He gas pressures used. The
various anion and cation currents were measured in the MCD upper cup
by selecting the appropriate polarities for the voltage on the LCD and
setting the MCD voltages to zero.

From conservation of particle flux, we expect at a given He pressure
$p$ to have
\begin{equation}
I_{\rm C^-}(p=0) = I_{\rm C^-}(p) + I_{\rm C}(p) + I_{\rm C^+}(p),
\end{equation}
where the subscripts label the beam currents, which are defined as
positive quantities.  This can be rewritten in terms of measured
quantities as
\begin{equation}
\frac{I_{\rm C^-}^{\rm u}(p)}{T_{\rm u}} 
= \frac{I_{\rm C^-}^{\rm u}(0)}{T_{\rm u}} 
- \frac{I_{\rm ND}(p)}{\gamma T_{\rm n}} 
- \frac{I^{\rm u}_{\rm C^+}(p)}{T_{\rm u}},
\end{equation}
where $I^{\rm u}$ are the currents measured in the upper cup and the
other quantities have been defined previously.  Due to the
configuration of the LCD, either $I_{\rm C^-}^{\rm u}$ or $I_{\rm
  C^+}^{\rm u}$ can be measured simultaneously with $I_{\rm ND}$, but
not both.  So it is useful to rearrange this equation as
\begin{equation}
I_{\rm C^-}^{\rm u}(p)
= I_{\rm C^-}^{\rm u}(0)
- I_{\rm ND}(p)
\left[\frac{{T_{\rm u}}} {\gamma T_{\rm n}}
+ \frac{I^{\rm u}_{\rm C^+}(p)}{I_{\rm ND}(p)}
\right].
\end{equation}
In order to determine $\gamma$, the currents $I_{\rm C^-}^{\rm u}$ and
$I_{\rm ND}$ are measured simultaneously for a set of pressure values
$p_i$.  The same is done for the currents $I_{\rm C^+}^{\rm u}$ and
$I_{\rm ND}$ but as a result of the coarse control of the leak valve
into the chicane, these data are collected on a different grid of
pressure values $p_k$.  Each of these latter two currents are then fit
to polynomials $P_{\rm C^+}(p_k)$ and $P_{\rm ND}(p_k)$, respectively,
which allows us to interpolate the data onto the $p_i$ pressure grid.
Expressing the ratio $I^{\rm u}_{\rm C^+}(p)/I_{\rm ND}(p)$ as the
ratio of these polynomials then yields
\begin{equation}
I_{\rm C^-}^{\rm u}(p_i)
= I_{\rm C^-}^{\rm u}(0)
- I_{\rm ND}(p_i)
\left[\frac{T_{\rm u}} {\gamma T_{\rm n}} 
+ \frac{P_{\rm C^+}(p_i)}{P_{\rm ND}(p_i)}
\right].
\end{equation}
Using this equation, we perform a least squares fitting over the
entire pressure range for the measured $I_{\rm C^-}^{\rm u}$ and
$I_{\rm ND}$ data and treat the quantities $I_{\rm C^-}^{\rm u}(0)$
and $\gamma$ as fitting parameters.  Doing this we find $\gamma = 2.7
\pm 0.3$ where the $\approx 11\%$ uncertainty represents the
run-to-run $1\sigma$ spread in the measured $\gamma$ and is treated as
a systematic uncertainty.

\subsection{CEM Efficiency}
\label{sec:cemefficiency}

We used a commercially available CEM to detect the ions resulting from
the reactions studied. CEMs have been shown to have a detection
efficiency of $\gtrsim 95\%$ for cations with incident kinetic
energies above 2~keV~amu$^{-1}$ \citep{Cran75a,Savi95a}. To better
quantify this, we have measured the detection efficiency of a CEM
identical to that used for our chemical studies here. For this we used
a beam of $\sim 27$~keV C$^+$ ($\sim 2.25$~keV amu$^{-1}$) as a proxy
for the CH$_n^+$ product ions ($\sim 2.33$~keV amu$^{-1}$). These
measurements were performed on a merged-beams apparatus at the
Universit\'e catholique de Louvain. The apparatus and technique is
described briefly below. A more technical description of the general
apparatus used for calibration can be found in \citet{Casa04a}.

Starting with an electron cyclotron resonance (ECR) ion source, we
extracted cations from the source and mass selected them to form a
25~keV beam of C$^{++}$. The beam was then electrostatically deflected
into a high vacuum region and into a floating cell. This deflection
prevented any C$^{+}$ formed upstream from making it into the floating
cell, as such ions had the wrong energy-to-charge ratio to be
transmitted into the cell. The floating cell was held at potential of
$-2$~kV, thus C$^{++}$ ions entering the floating cell were
accelerated to 29~keV.  Inside the floating cell, a fraction of the
C$^{++}$ beam underwent electron capture to form 29~keV C$^+$.
C$^{++}$ ions exiting the floating cell decelerated to 25~keV and the
C$^+$ to 27~keV.
 
After the floating cell, both carbon charge states are then separated
using an electro-magnet which directs the $\sim$~nA 25~keV C$^{++}$
current into a movable Faraday cup within the magnet and the $\sim$~pA
27~keV C$^+$ current, generated in the floating cell, into a
retractable Faraday cup at the exit of the magnet. The floating cell
voltage shifts the energy-to-charge ratio, so that only those ions
formed via electronic capture within the floating cell have the
correct energy-to-charge ratio to reach the detector. Any C$^+$
current produced outside of the floating cell will have a
  kinetic energy of 25~keV and not be directed into the retractable
Faraday cup.  Comparing the two currents we found that $\approx 7.5
\times 10^{-4}$ of the C$^{++}$ underwent electron capture to form
C$^+$. We then attenuated the C$^{++}$ beam at the source to give an
$\sim$~pA C$^{++}$ current as measured within the magnet.  The changes
in the source conditions had no measurable effect on the pressure in
the floating cell or in the magnet.  Hence, as we are in the single
collision regime, it is safe to assume that the attenuated C$^+$
particle count rate should be given by $\approx (7.5 \times
10^{-4})I_{\rm C^{++}}/e$.  Comparing this predicted value to that
measured on a CEM situated directly behind the retractable Faraday
cup, we find the CEM efficiency to be $\eta=0.99 \pm 0.03$.

\section{Beam Overlap and Relative Energies}
\label{sec:energies}

We have determined the overlap of the two beams through a combination
of beam profile measurements and geometric modeling. The beam profiles
are measured using the two BPMs located in the interaction
region. Typical profiles are shown in Figure~\ref{Fig:Profiles}.
Simulations of the beam trajectories have been created based on the
known experimental geometry and the measured beam profiles. These
modeling studies are also used to determine the average relative
energy $\langle E_{\rm r} \rangle$ and the corresponding energy
spread. The methodology employed is similar to that described by
\citet{Bruh10a} and is only briefly reviewed here.

\subsection{Beam overlap}
\label{sec:Over}

The overlap between the two beams at an arbitrary position is given by
\begin{equation}
\label{eq:overlap}
{\Omega(z) = \frac{\iint{J_\mathrm{n}(x,y,z)J_\mathrm{i}(x,y,z)\,dx\,dy}}
{\iint{J_\mathrm{n}(x,y,z)\,dx\,dy}\iint{J_\mathrm{i}(x,y,z)\,dx\,dy}}},
\end{equation}
where $J_{\rm n}$ and $J_{\rm i}$ are the fluxes of the neutral and
ion beams, respectively; $z$ is chosen to lie along an axis defined by
the propagating beams in the laboratory frame; and $x$ and $y$ are
both perpendicular to the $z$ axis and to one another.
\citet{Bruh10a} explain how this is implemented experimentally using
the BPM data to calculate $\Omega(z)$. These experimental values are
used to constrain the geometric simulations described below. The
geometric model enables us, in turn, to determine the average overlap
factor in the interaction region
\begin{equation}
\left\langle\Omega(z)\right\rangle = \frac{1}{L}\int_0^L{\Omega(z)}\, dz,
\end{equation}
which is needed for Equation~(\ref{eq:rate}). 

\subsection{Geometric Simulations}
\label{subsec:Geometric}

Particle ray tracing was computed starting from a pseudo-plane
analogous to the end of the interaction region. In this Monte Carlo
simulation, the particles were flown in reverse from this plane and
through the limiting apertures of their respective beam lines. Each
particle was given a random starting position and an initial
trajectory in this pseudo-plane. In addition, the kinetic
energy for each particle was randomly assigned from a Gaussian
probability distribution using the FWHM for the corresponding source.

\subsubsection{Beam Profiles}
\label{subsubsec:Beam}

Beam profiles were calculated from the simulated particle flux
distribution at pseudo-plane analogs of the two BPM locations within
the interaction region (see Figure~\ref{Fig:Profiles}). The profiles
were derived along the lines of Equations~(53) and (54) of
\citet{Bruh10a}. The simulated C beam profiles were found to be in
good agreement with those measured. For the H$_3^+$ beam, it was
necessary to adjust the shape and position of the limiting apertures
used in the calculations in order for the simulated profiles to best
match those measured. These adjustments were required to account for a
vertical shift of the H$_3^+$ beam as it moved through the interaction
region along with focusing effects induced by the beam merger,
resulting in a typical average bulk misalignment of 0.81~mrad as
determined from the beam profiles.  SIMION studies indicate that these
features are the result of a minor misalignment of the 1D lens prior
to the 90$^\circ$ deflection which merges the H$_3^+$ onto the C beam.

From our Monte Carlo trajectory simulation, we can readily calculate
$\Omega(z)$ using Equation~(\ref{eq:overlap}). This is highly
advantageous as it is not possible to measure the beam profiles along
the entire interaction region. We use these simulations as a guide for
interpolating $\Omega(z)$ between the BPMs and for extrapolating
$\Omega(z)$ upstream and downstream of the BPMs.

\subsubsection{Relative Energies}
\label{subsubsec:collide}

Calculating the particle relative energies from
the Monte Carlo simulation required that the interaction region be
discretized into pseudo-planes. In turn, each pseudo-plane was further
divided into square cells. The size of the cells was selected so as to
ensure statistically significant particle densities. For every
neutral-ion pair within each cell, we calculated the interaction angle and relative velocity. This process was repeated
for every pseudo-plane. A typical simulation used 10,000 particles for
each species with over 50 pseudo-planes, each divided into 900 cells.
From these simulations, which take into account the bulk misalignment
of the beam and their angular spreads, and binning the resulting
interaction angles into a histogram, a Gaussian-like distribution
emerged yielding a mean interaction angle of $\left\langle \theta
\right\rangle= 1.16 \pm 0.46$~mrad.

\begin{figure}[!p]
\centering
\includegraphics[width=1\textwidth]{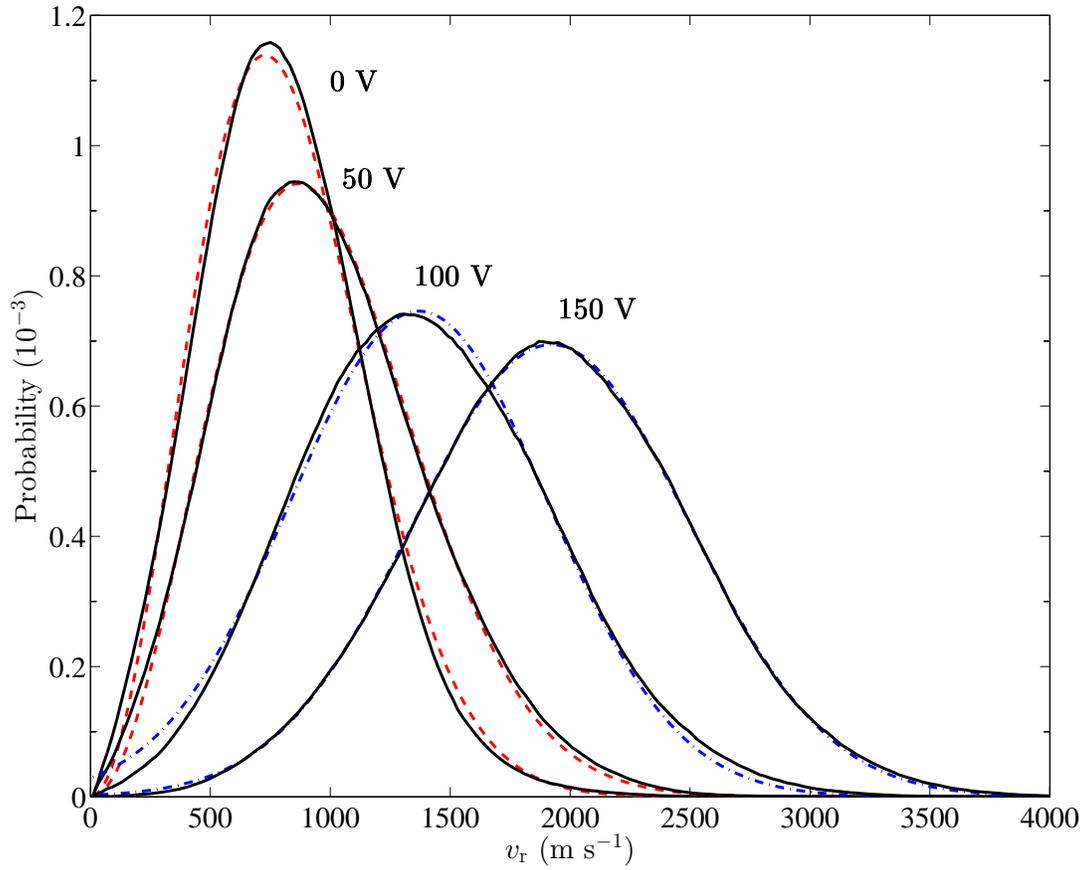} 
\caption{Relative velocity distribution for selected values of the
  floating cell voltage $|U_{\rm f}|$. The solid curves show the
  results for $|U_{\rm f}| =$ 0, 50, 100, and 150~V. The dashed curves
  are $\approx72$ and $109$~K Maxwell-Boltzmann distributions which
  provided the best fits to the 0 and 50~V, data respectively. The
  Gaussian fits of the 100 and 150~V results (dotted curves) are also
  shown.}
\label{Fig:VelProfile}
\end{figure}

\begin{figure}[!p]
\centering
\includegraphics[width=1\textwidth]{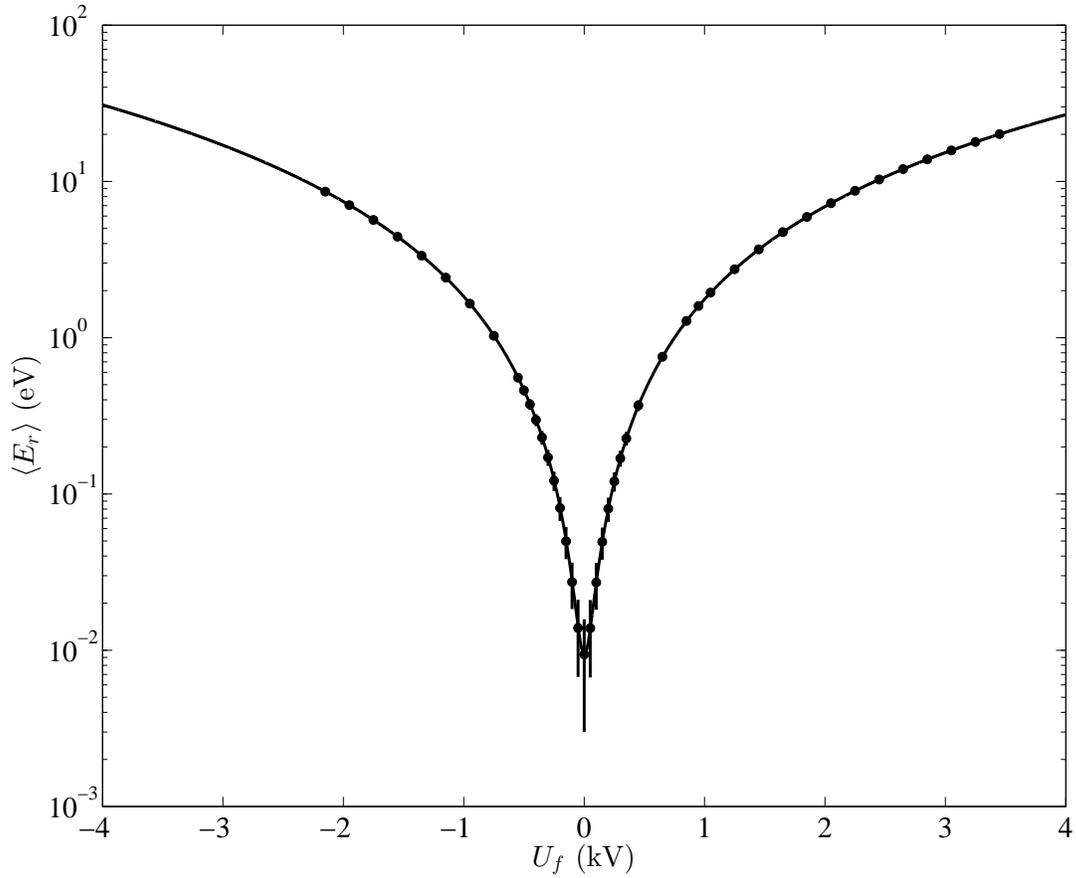} 
\caption{Simulated average relative energy
  $\langle E_r\rangle$ as a function of the floating cell voltage
  $U_{\rm f}$.  Vertical error bars on the filled circles indicate the
  FWHM spread of the modeled distribution. The solid line is the
  calculation for the average relative energy derived from
  Equation~(\ref{Eqn.brou}).}
\label{Fig:EVs.float}
\end{figure}

A histogram for the calculated values of $v_{\rm r}$ throughout the
interaction region is shown in Figure~\ref{Fig:VelProfile} for a small
sample of floating cell voltages.  For nearly velocity-matched beams
$|U_{\rm f}|\lesssim 50$~V, the relative velocity spread is
dominated by the bulk interaction angle and the divergence of the two
beams relative to one another. In this regime, the relative
velocity spread is well described by a Maxwell-Boltzmann
distribution\footnote{More precisely we used the form for a
  three-dimensional Maxwell-Boltzmann velocity distribution as
  expressed in spherical coordinates and then integrated over $\theta$
  and $\phi$, leaving the relative velocity $v_{\rm r}$ as the sole
  remaining variable.}.  For floating cell values of $|U_{\rm f}|
\gtrsim 100$~V the relative velocity spread is determined
largely by the energy spreads of the two ion sources.  The resulting
function tends towards a Gaussian distribution in velocity.  The
simulations also enable us to determine $\left\langle E_{\rm r}
\right\rangle$ versus $U_{\rm f}$, along with the corresponding energy
spread (see Figure~\ref{Fig:EVs.float}). The simulations of the ${\rm
  C + H_3^+}$ reaction complex studied here indicate that we achieved
values of $\left\langle E_{\rm r} \right\rangle$ as low as
$\approx 9.3$~meV, corresponding to an effective translational
temperature of $\approx 72$~K (as derived from a Maxwell-Boltzmann fit
of the velocity distribution).

\section{Signal Determination}
\label{sec:Signal}

\begin{figure}[!p]
\begin{center}
\includegraphics[width=1\textwidth]{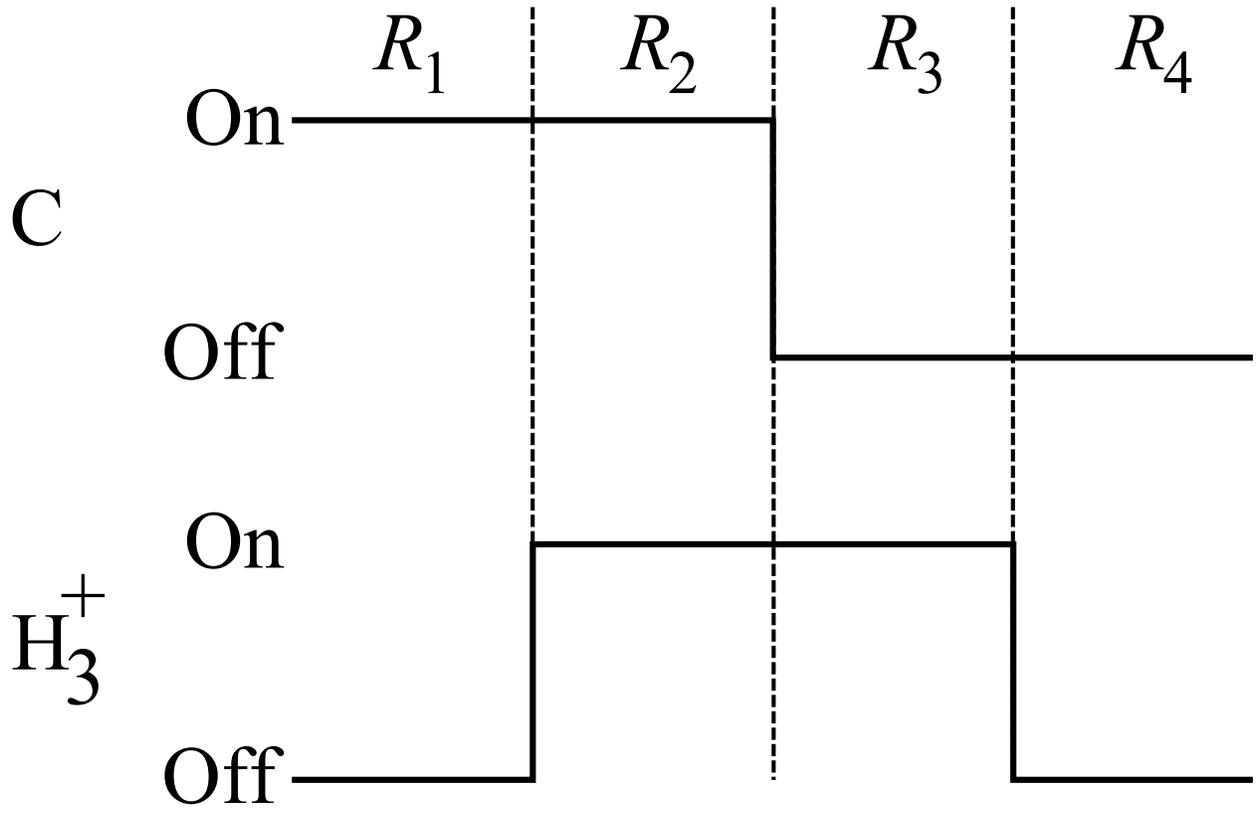} 
\caption{Data acquisition timing sequence.}
\label{Fig:timing}
\end{center}
\end{figure}

The signal rate was extracted by accounting for the various
backgrounds using a standard beam chopping technique \citep{Brou83a}.
The neutral beam is chopped by turning the laser on and off.  The
timing of the laser is controlled using a programmable digital signal
from the power supply unit (PSU).  During typical operation the laser
was gated on for 5~ms and off for 5~ms.  The PSU also provides an
external trigger.  We sent this through a gate-and-delay generator and
fed it into the fast high voltage switch that we use to chop the
H$_3^+$ beam.  The delay time is set to 2.5~ms or a quarter of the
period for the laser chopping pattern.  The resulting square wave
pattern used is shown in Figure~\ref{Fig:timing}.

The CEM counts for each quadrant of the chopping cycle are recorded in
four individual counter channels as
\begin{align}
N_1 & =           N_{\rm C}              + N_{\rm b},  \label{Eqn.N1}\\
N_2 & = N_S     + N_{\rm C} + N_{\rm H_3^+} + N_{\rm b},  \label{Eqn.N2}\\
N_3 & =                     N_{\rm H_3^+} + N_{\rm b},  \label{Eqn.N3}\\
N_4 & =                                  N_{\rm b}. \label{Eqn.N4}
\end{align}
Here $N_S$ represents the signal counts with both beams on, $N_{\rm
  C}$ is the background counts due to the C beam, $N_{\rm H_3^+}$
is the background counts due to the H$_3^+$ beam, and $N_{\rm b}$
is the background counts with both beams off.  The corresponding
uncertainty in $N_S$ from counting statistics is given by
\begin{equation}
\delta N_S = \left(N_1 + N_2 + N_3 + N_4\right)^{1/2}.
\label{eq:deltaNS}
\end{equation}
Data are collected at a given relative energy,
for an integration time $t$.  Taking the chopping pattern into
account, the corresponding counter rates are readily calculated by
dividing $N_i$ by $t/4$, yielding $R_i$.  The signal rate is then
given by
\begin{equation}
S = R_2 - R_1 - R_3 + R_4 \label{Eqn.S2}
\end{equation}
where the fractional uncertainty in $S$ is $\delta N_S/N_S$.

\begin{figure}[!p]
\centering
\includegraphics[width=1\textwidth]{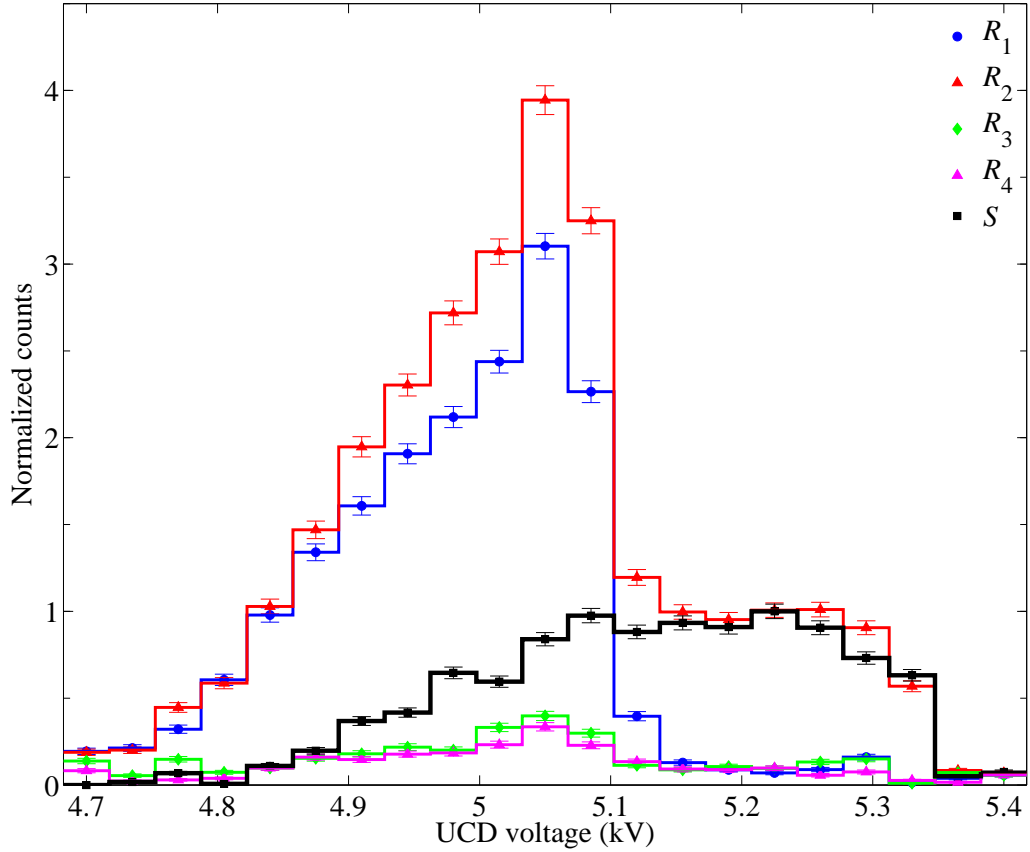} 
\caption{UCD scans of rates $R_1$ through $R_4$ and the resulting
  extracted signal $S$. The data have been normalized so that the peak
  in $S$ is 1. The error bars show the $1\sigma$ counting-statistics
  uncertainty on each point.} 
\label{Fig:UCDChopScan}
\end{figure}

Figure~\ref{Fig:UCDChopScan} shows count rates $R_1$ through $R_4$ as
a function of the UCD plate potential.  For this the LCD and MCD
voltages were already optimized for transmittance of the CH$^+$
signal.  These rates have been normalized to the peak of the resulting
signal rate scan, which is also shown in the figure.  The largest of
these beam-induced backgrounds occurs when the C beam is on.  This
background is due to C atoms stripping on residual gas and forming
C$^+$ ions with a kinetic energy close to that of the
CH$^+$ signal ions.  A portion of the resulting C$^+$ ions are
transmitted through the electrostatic energy analyzer system and into
the CEM for UCD voltages just below those giving the optimal
transmission of the CH$^+$ beam.

\begin{figure}[!p]
\centering
\includegraphics[width=1\textwidth]{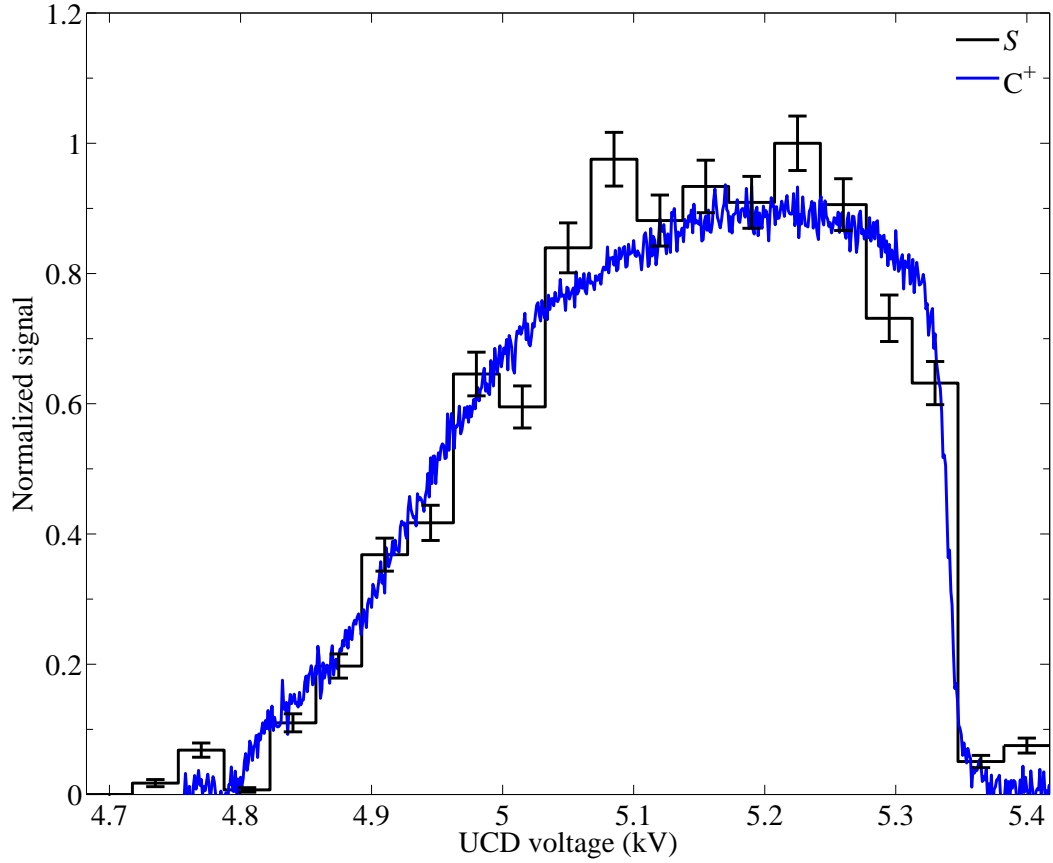} 
\caption{UCD voltage scan of a C$^+$ beam current, with a
    kinetic energy tuned to act as a proxy for the CH$^+$
  signal.  Also shown is the measured signal count rate $S$ and
  associated $1\sigma$ counting-statistics uncertainty. $S$ has been
  normalized to 1 at the peak value and the C$^+$ current has been
  scaled to best show the agreement in the structure between the two.}
\label{Fig:UCDScan}
\end{figure}

We confirmed the shape of the resulting signal scan using a proxy for
the CH$^+$ signal.  The proxy was generated by tuning the kinetic
  energy of the C$^-$ beam to that expected for the CH$^+$ signal.
Double electron detachment on residual gas in the apparatus generated
a C$^+$ beam which we directed through the detector system and into a
Faraday cup mounted at the CEM position.  A comparison of the
normalized proxy beam current is shown in Figure~\ref{Fig:UCDScan},
along with the normalized signal counts.  The transmitted profiles
match closely, confirming that we have successfully removed the
background from the signal.  The proxy C$^+$ scan profile skews
similar to the signal profile.  We carried out experimental tests
which confirmed that this is due to the beams entering the LCD from
the chicane at a slight angle and a mismatch in applied plate voltages
with respect to that required for ion transmission along the central
trajectory of the cylindrical deflectors.

\section{Data Acquisition Procedure}
\label{sec:DAQ}

For a typical data acquisition cycle, the C$^-$ and H$_3^+$ beams are
first each tuned independently to optimize the transported current and
beam profiles in the interaction region.  The beams are then tuned
together to make them as parallel as possible.  Once tuned, data
acquisition begins.  The data acquisition procedure is largely
controlled via an automated \textsc{L}ab\textsc{view} program.  For a
typical data acquisition run, $U_\mathrm{f}$ is fixed and the signal
counts $N_S$ recorded until the statistical uncertainty $\delta
N_S/N_S$ approaches $\sim 4\%$.

The floating cell voltage was scanned in steps of 50~V for $|U_{\rm
  f}|$ between 0 and $500$~V and from there in larger steps of 100~V
up to $2.0$~kV.  In order to minimize focusing of the C$^-$ beam by
the floating cell, and hence of the C beam, we limited $|U_{\rm f}|$
to voltages below 10\% of the sputter source voltage, typically
$|U_{\rm f}|\leq$ 2~keV.  To achieve effective values of
$|U_\mathrm{f}|$ beyond $2.0$~kV, the C$^-$ ion source voltage was
offset and $U_\mathrm{f}$ scanned in 200~V steps from 0 to 2.0~kV.
Varying the voltage offset on the source enabled us to achieve
effective values of $U_{\rm f}$ up to $4.0$~kV. The upper limit was
defined by the maximum stable source potential of $\approx 30$~kV,
combined with $U_{\rm f}$ = 2~kV.  As a result, we were able to scan
$\left\langle E_{\rm r} \right\rangle$ between
$\approx 9$~meV and 20~eV.

Data acquisition for a typical data run begins with the control
program measuring the C and H$_3^+$ beam profiles.  While one beam is
being profiled, the other is off.  Beam chopping is then initiated and
the beam currents measured.  To within the stability of each beam, the
chopped current measurement is half that of the unchopped beam.  Next,
all four counters are initialized and data are collected for an
integration time of $t = 10$~s.  The current measurements and 10-s
integration are repeated typically $100-200$ times.  Afterwards, the
chopping is turned off and the beam profiles are measured again.
During the period between profile measurements, the beam currents are
continually monitored and act as a proxy for the stability of the
measurement and the alignment of the beams.  The data acquisition
cycle is repeated until either the statistical uncertainty approaches
the desired level or the ion beams begin to de-tune.

\section{Uncertainties and Averaging}
\label{sec:Uncert}

Tables~\ref{Tab:Values1} and \ref{Tab:Values2} list typical
experimental operating values for the quantities going into
Equation~(\ref{eq:rate}) and their associated uncertainties.  The
different terms are divided into those uncertainties which are
statistical in nature (Table~\ref{Tab:Values1}) and those which are
systematic (Table~\ref{Tab:Values2}).  The signal data at a given
relative energy were collected over a number of experimental runs
until the counting-statistics uncertainty in $S$ was typically less
than 4\%.  More details about the various terms can be found in
Sections~\ref{sec:Apparatus}-\ref{sec:Signal}.

For each data run $i$ at a given $\left\langle E_{\rm r}
\right\rangle$, we calculated the measured merged-beams rate
coefficient $\left\langle\sigma v_\mathrm{r}\right\rangle_i$ and the
associated statistical-like experimental uncertainty
$\delta\left\langle\sigma v_\mathrm{r}\right\rangle_i$.  The various
runs at that energy were averaged together using a weighting of
\begin{equation}
w_i = \frac{1}{(\delta\left\langle\sigma v_r\right\rangle_i)^2}
\end{equation}
The resulting merged-beams rate coefficient and associated $1\sigma$
statistical-like uncertainty is then given by
\begin{equation}
\left\langle\sigma v_r\right\rangle 
= \frac{\sum_i{\left\langle\sigma v_r\right\rangle_i w_i}}{\sum_i{w_i}} 
\pm \left(\sum_i{w_i}\right)^{-1/2}.
\end{equation}
There is an additional 12\% systematic uncertainty on each resulting
merged-beams rate coefficient.

\section{Results}
\label{sec:results}

\subsection{Experimental Merged-Beams Rate Coefficients}
\label{sec:ExpRate}

\begin{figure}[!p]
\centering
\includegraphics[width=1\textwidth]{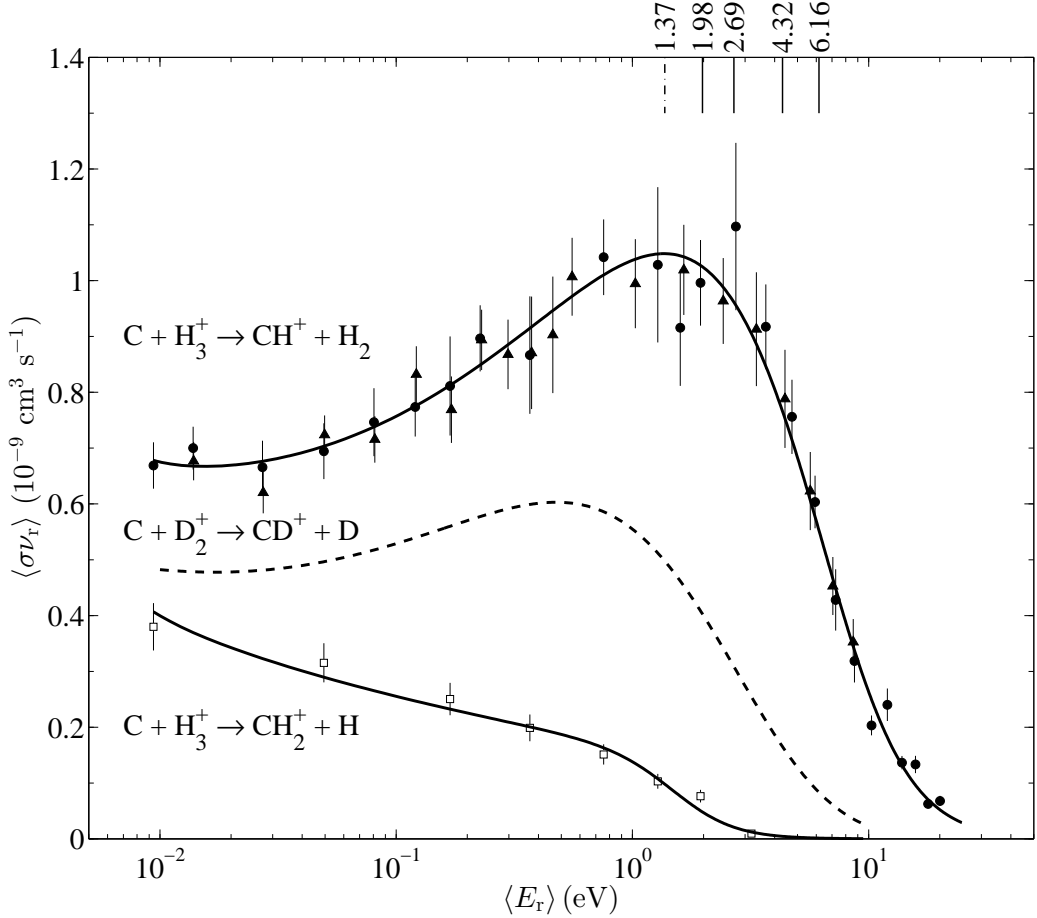} 
\caption{Experimental merged-beams rate coefficient
  $\left\langle\sigma v_r\right\rangle$ as a function of the average
  relative energy $\left\langle E_r\right\rangle$
  for reaction~(\ref{eq:CH3+_pT}) forming $\mathrm{CH}^{+} +
  \mathrm{H}_{2}$ (filled symbols) and reaction~(\ref{eq:CH3+_ppT})
  forming $\mathrm{CH}_2^+ + \mathrm{H}$ (open symbols). The circles
  denote $v_{\rm C} \geq v_{\rm H_3^+}$ and the triangles $v_{\rm C} <
  v_{\rm H_3^+}$ for reaction~(\ref{eq:CH3+_pT}). The error bars
  signify the 1$\sigma$ statistical-like uncertainty. The solid lines
  are an empirical fit to the experimental data using
  Equation~(\ref{Eqn.rateCH}). For comparison, the dashed line shows
  the measured experimental rate coefficient of
  \protect\citet{Schu83a} for reaction~(\ref{eq:CD2+_DT}): $\mathrm{C}
  + \mathrm{D}_2^+ \rightarrow \mathrm{CD}^+ + \mathrm{D}$. The solid
  vertical lines denote energies at which the competing
  reactions~(\ref{eq:Diss1}), (\ref{eq:CH}), (\ref{eq:Diss2}), and
  (\ref{eq:Diss3}) open at $\approx 1.98$, $2,69$, $4.32$, and
  $6.16$~eV, respectively. The dot-dashed vertical line denotes the
  energy at which the experimental rate coefficient is inferred to
  peak, approximately $0.61$~eV below the first competing channel.}
\label{Fig:ExpRate}
\end{figure}

Figure~\ref{Fig:ExpRate} presents our experimental results for the
merged-beams rate coefficient as a function of the average 
  relative energy $\langle E_{\rm r} \rangle$ for
$\mathrm{C+H_3^+\rightarrow CH^+ + H_2}$, reaction~(\ref{eq:CH3+_pT}),
and for $\mathrm{C + H_3^+ \rightarrow CH_2^+ + H}$,
reaction~(\ref{eq:CH3+_ppT}).  The error bars show the $1\sigma$
statistical-like uncertainty.  We also searched for CH$_3^+$ signal
from reaction~(\ref{eq:CH3+_RA}): $\mathrm{C+H_3^+\rightarrow CH_3^+ +
  photon}$.  However, analyzer scans within the predicted 
  kinetic energy range for the CH$_3^+$ signal yielded count rates
indistinguishable from the background noise. At matched beam
velocities, $\langle E_{\rm r} \rangle = 9.3$~meV, the measured rate
coefficient of $-0.657 \pm 6.42 \times 10^{-11}$~cm$^3$~s$^{-1}$,
enables us to put a $1\sigma$ upper limit of $5.76 \times
10^{-11}$~cm$^3$~s$^{-1}$ on this channel.

\subsection{Cross Sections}
\label{Sec.crosssection}

We have extracted the cross section from our data using the functional
form
\begin{equation}
\sigma_{x} = \frac{a_0+a_{1/2}E^{1/2}}{E^{2/3}+b_1E+b_2E^2+b_4E^4},
\label{Eqn.rateCH}
\end{equation}
where ${x}$ denotes either reaction~(\ref{eq:CH3+_pT}) or
(\ref{eq:CH3+_ppT}). The resulting cross sections are in units of
cm$^2$ for $E$ in eV.  Over the ranges for which data were measured,
the fitting accuracy was between $2-6$\% for
reaction~(\ref{eq:CH3+_pT}) and $6-17$\% for
reaction~(\ref{eq:CH3+_ppT}).

Concerning the first term of the denominator in
  Equation~(\ref{Eqn.rateCH}), leaving the power as a free fitting
  parameter yields an $E^{-0.7 \pm 0.1}$ behavior at low energies for
  CH$^+$ formation.  For the case of CH$_2^+$, we find an $E^{-0.3 \pm
    0.3}$ behavior.  We attribute the large uncertainty in this term
  to the rapid decrease of the CH$_2^+$ merged-beams rate coefficient
  with relative energy.  As a result the fit is dominated by the
  higher order terms.  Here, we have chosen to use $E^{-2/3}$ for both
  channels as it agrees to within the experimental uncertainties and
  for which there is some theoretical support. This term results in a
thermal rate coefficient with a $T^{-1/6}$ behavior at low
temperatures and matches the calculated behavior for the thermal
rate coefficient of the electronically similar reaction complex
$\mathrm{O}(^3P) + \mathrm{H_3^+}$, which is predicted to be dominated
at low temperatures by the charge-quadrupole interaction
\citep{Klip10a}.  The terms in the denominator with greater powers of
$E$ have been arbitrarily selected to match the higher energy
dependence in each of the measured merged-beams rate coefficients.

\begin{figure}[!p]
\centering
\includegraphics[width=1\textwidth]{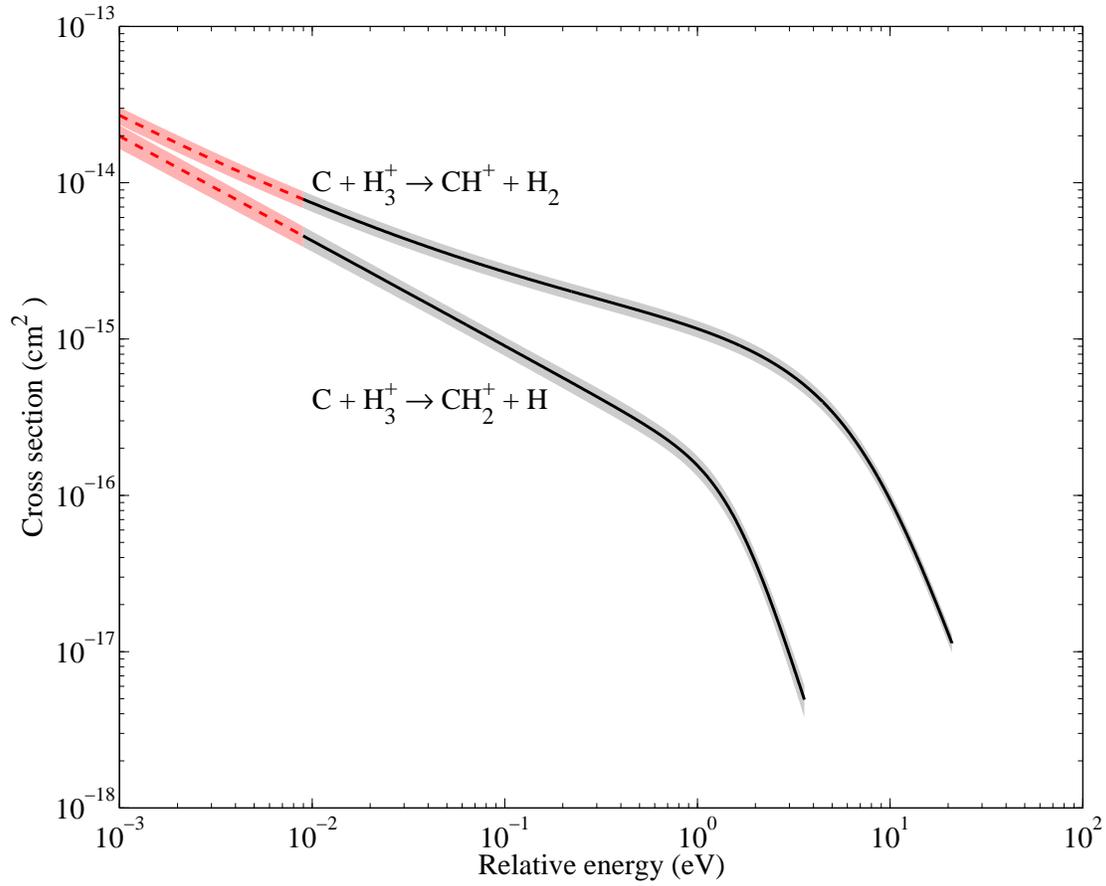} 
\caption{Experimentally derived cross sections as a function of 
    relative energy for reactions~(\ref{eq:CH3+_pT}) and
  (\ref{eq:CH3+_ppT}) are shown by the solid black lines. The shaded
  areas signify the quadrature sum of the systematic uncertainty and
  fitting accuracy. The red lines use the fits to extrapolate the
  experimental results to lower impact energies and the surrounding
  shaded region assumes a constant uncertainty given by that at the
  lowest measured energy.}
\label{Fig:Cross-section}
\end{figure}

The best fit parameters were derived using these functional forms for
the cross sections, multiplying them by $v_{\rm r}$, convolving them
with the experimental velocity distribution, and performing a $\chi^2$
fit between the measured merged-beams rate coefficients and the model.
The resulting best fits to the data are shown by the solid lines in
Figure~\ref{Fig:ExpRate}.  For reaction~(\ref{eq:CH3+_pT}) the fit is
good over the measured relative energy range of $\approx 9$~meV
to 20~eV and for reaction~(\ref{eq:CH3+_ppT}) from $\approx 9$~meV to
3~eV.  The best fit parameters of the cross section for each reaction
are given in Table~\ref{tab:rate_CH}. The experimentally derived cross
sections for both reactions are plotted in
Figure~\ref{Fig:Cross-section}.

\subsection{Translational Temperature Rate Coefficients}
\label{sec:kineticT}

The translational temperature rate coefficient $\alpha_x$, for
reaction $x$, is derived by multiplying the extracted cross section
$\sigma_x$ by the relative velocity and convolving the product with a
Maxwell-Boltzmann distribution.  Using Equation~(\ref{Eqn.rateCH}) as
a guide, we have fit our resulting rate coefficients with
\begin{equation}
\alpha_{x} = 
\frac{a_0+a_{1/2}T^{1/2}+a_1T}{T^{1/6}+b_{1/2}T^{1/2}+b_1T+b_{3/2}T^{3/2}},
\label{Eqn.thermalrateCH} 
\end{equation}
where ${x}$ denotes either reaction~(\ref{eq:CH3+_pT}) or
(\ref{eq:CH3+_ppT}). The resulting rate coefficients are given in
units of cm$^3$~s$^{-1}$ for $T$ in units of K.  The best fit
parameters for each reaction are given in Table~\ref{tab:thermal_CH}.

\begin{figure}[!p]
\centering
\includegraphics[width=1\textwidth]{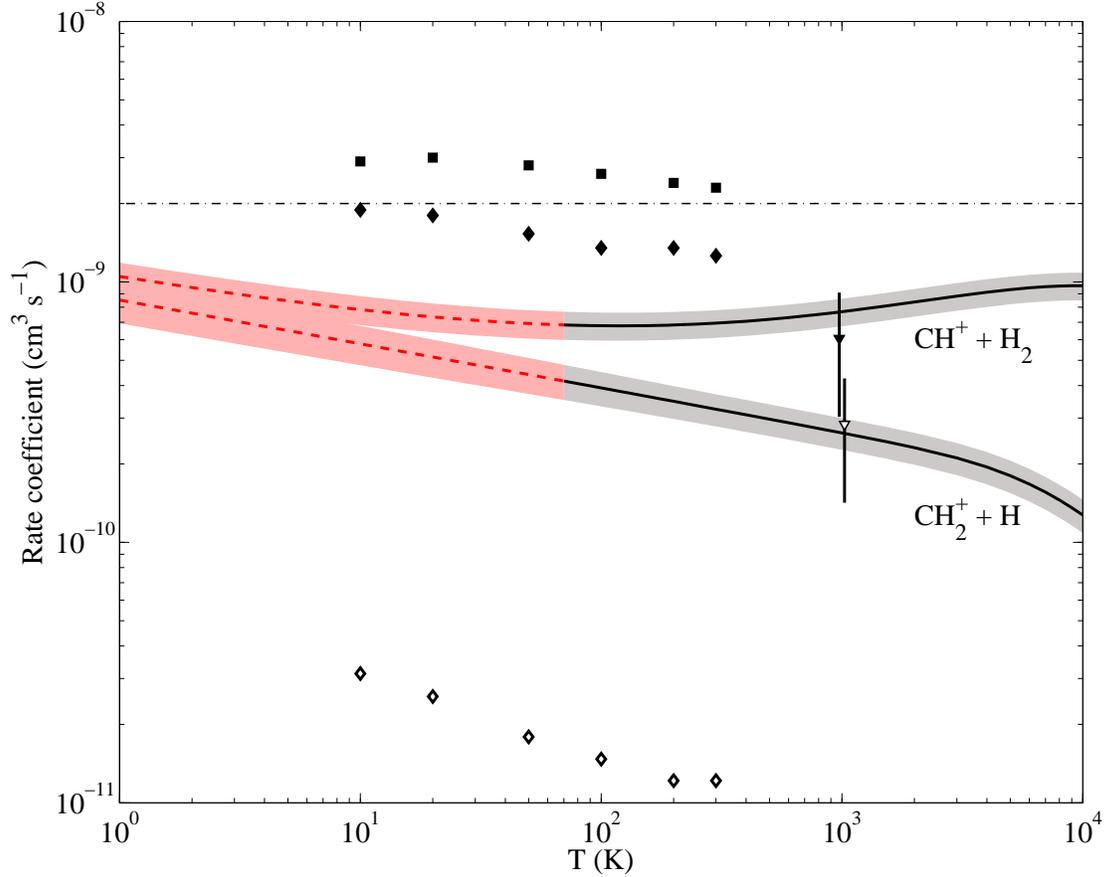} 
\caption{The black solid lines present our experimentally derived translational temperature rate coefficient for $\mathrm{C} +
  \mathrm{H_3^+} \rightarrow \mathrm{CH}^+ + \mathrm{H}_2$,
  reaction~(\ref{eq:CH3+_pT}), and $\mathrm{C} + \mathrm{H_3^+}
  \rightarrow \mathrm{CH_2}^+ + \mathrm{H}$,
  reaction~(\ref{eq:CH3+_ppT}).  The quadrature sum of the systematic
  uncertainty and fitting accuracy is denoted by the shaded region.
  The red dashed lines extrapolate these results to lower temperatures
  and the surrounding shaded area assumes a constant systematic
  uncertainty given by that at the lowest measured temperature added
  in quadrature to the accuracy of the fit.  The dot-dashed curve
  shows the Langevin rate coefficient.  The theoretical thermal rate
  coefficients of \protect\citet{Talb91a} and
  \protect\citet{Bett98a,Bett01a} are shown by the squares and
  diamonds, respectively. The full and open symbols denote the results
  for reactions~(\ref{eq:CH3+_pT}) and (\ref{eq:CH3+_ppT}),
  respectively.  The inverted triangles give the experimental result
  of \protect\citet{Savi05a} for the fully deuterated isotopologues
  for these reactions, but scaled by the reduced mass for $\mathrm{C +
    H_3^+}$ collision system. Their results are at an estimated translational temperature of $\sim 1,000$~K and for clarity have
  been shifted by $\mp 25$~K, respectively.}
\label{Fig:thermalrate}
\end{figure}

The experimentally derived translational temperature rate
coefficients are shown by the solid curves in
Figure~\ref{Fig:thermalrate}.  The shaded regions show the quadrature
sum of the systematic uncertainty and the fitting accuracy, yielding
an uncertainty of between $12-13$\% for reaction~(\ref{eq:CH3+_pT})
and between $14-18$\% for reaction~(\ref{eq:CH3+_ppT}).  The low
temperature limit for the validity of the derived translational
temperature rate coefficients is $\approx 72$~K, which is the
effective translational temperature of our experimental energy
spread for the minimum $\langle E_{\rm r} \rangle$ achieved. The
functional form of Equation~(\ref{Eqn.thermalrateCH}) has been chosen
so that the extrapolation below 72~K goes to a $T^{-1/6}$ behavior as
predicted by \citet{Klip10a} for the electronically similar reaction
complex $\mathrm{O}(^3P) + \mathrm{H_3^+}$.  The high temperature
limits for the fits of $\approx 2.3 \times 10^5$ and $3.5 \times
10^4$~K correspond to the highest values of $\langle E_{\rm r}
\rangle$ measured for reactions~(\ref{eq:CH3+_pT}) and
(\ref{eq:CH3+_ppT}) of $\approx 20$ and 3~eV, where the cross sections
are vanishingly small.

\section{Discussion}
\label{sec:discussion}

\begin{figure}[!p]
\centering
\includegraphics[width=1\textwidth]{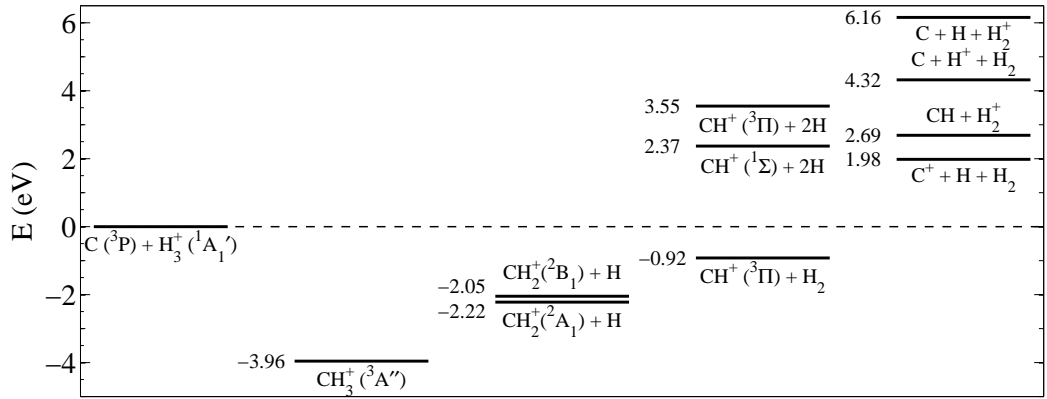} 
\caption{Energy-level diagram for various C + H$_3^+$ reaction
  pathways, as given by \protect\citet{Dela14a}}.
\label{Fig:Energybalance}
\end{figure}

An energy-level diagram for the various ${\rm C + H_3^+}$ reaction
pathways which we discuss in this section can be found in
Figure~\ref{Fig:Energybalance}. The sources for the derivation of this
energy-level diagram are given in the figure caption.

\subsection{Merged-Beams Rate Coefficient}

\subsubsection{$\mathrm{C+H_3^+ \rightarrow CH^+ + H_2}$}

The neutral C is of $^3P$ symmetry. H$_3^+$ is of $^1$A$^\prime_1$
symmetry, as this is the only electronic state which lies below the
dissociation limit of the molecule \citep{McNa95a}. The final
electronic state of H$_2$ is $^1\Sigma_g^+$.  Hence, taking the spin
multiplicities into account (i.e., ignoring possible intersystem
  transitions to the CH$_3^+$ singlet manifold), the lowest
accessible symmetry of CH$^+$ is the $a^3\Pi$ electronic state
\citep{Talb91a,Bett98a}.  Putting it all together we can re-write
reaction~(\ref{eq:CH3+_pT}) as
\begin{equation}
{\rm C}(^3P) + {\rm H}_3^+(^1{\rm A}^\prime_1) 
\rightarrow 
{\rm CH}^+(a^3\Pi) + {\rm H}_2(^1\Sigma_g^+).
\end{equation}
This reaction is exoergic by $\approx 0.92$~eV \citep{Dela14a}.  The
only additional channel for CH$^+$ formation is the endoergic
reaction:
\begin{equation}
\label{eq:CH3+_pTdH2}
{\rm C}(^3P) + {\rm H}_3^+(^1{\rm A}^\prime_1) 
\rightarrow 
{\rm CH}^+(a^3\Pi) + {\rm H}(^2S) + {\rm H}(^2S).
\end{equation}
The threshold for this reaction is $\approx 3.55$~eV.\cite{Dela14a}

Our measured merged-beams rate coefficient for
reaction~(\ref{eq:CH3+_pT}) exhibits a relative energy
dependence that is similar to that measured by \citet{Schu83a} for
reaction~(\ref{eq:CD2+_DT}): $\mathrm{C + D_2^+ \rightarrow CD^+ +
  D}$.  Figure~\ref{Fig:ExpRate} presents their cross section results
multiplied by $v_{\rm r}$.  Both measurements show an initial increase
in the merged-beams rate coefficient with increasing relative energy.
Possible reasons for the similar behaviors seen with increasing
  energy could be due to the opening up of new electronic states in
  the intermediate reaction complex or to additional ro-vibrational
channels becoming energetically accessible in the daughter products.
Clearly, though, further theoretical and experimental work is
  needed to understand the observed behavior.

At some point, the magnitude of the rate coefficient dramatically
decreases with increasing relative energy.  We attribute this to
the opening of additional reaction pathways that compete with the
reaction we are measuring.  The first four of these channels are:
\begin{eqnarray}
\mathrm{C} + \mathrm{H_3^+} 
&\rightarrow&
\mathrm{C}^+ + \mathrm{H} + \mathrm{H}_2, \label{eq:Diss1}\\
\mathrm{C} + \mathrm{H_3^+} 
&\rightarrow&
\mathrm{CH} + \mathrm{H}_2^+, \label{eq:CH}\\
\mathrm{C} + \mathrm{H_3^+} 
&\rightarrow&
\mathrm{C} + \mathrm{H}^+ + \mathrm{H}_2, \label{eq:Diss2}\\
\mathrm{C} + \mathrm{H_3^+} 
&\rightarrow&
\mathrm{C} + \mathrm{H} + \mathrm{H}_2^+, \label{eq:Diss3} 
\end{eqnarray}
with threshold energies of $\approx$ 1.98, 2.69, 4.32, and
  6.16~eV, respectively (see Figure~\ref{Fig:Energybalance}). In
this relative  energy range, we also see no obvious sign for the
onset of CH$^+$ formation via reaction~(\ref{eq:CH3+_pTdH2}).  We
attribute this, in part, to the opening of the above competing
channels.

A similar decrease was seen by \citet{Urba91a} for the associative
ionization (AI) reaction $\mathrm{H}(1s) + \mathrm{H}(2s) \rightarrow
\mathrm{H_2^+} + e^-$. One difference, though, is that the cross
section for the AI reaction shows a sharp and dramatic decrease at the
opening of the competing $\mathrm{H}(1s) + \mathrm{H}(2s) \rightarrow
\mathrm{H}(1s) + \mathrm{H}^+ + e^-$ channel. This is readily
explained by the well-defined initial internal energies of the
reactants and the absence of any internal degrees of freedom in their
products.

In our results we cannot unambiguously identify the opening of any of
the above channels competing with reaction~(\ref{eq:CH3+_pT}).
Moreover, the decrease seen in our data is not as sharp as that seen
by \citet{Urba91a}.  Both the shift and broadening of the observed
threshold are most likely due to the H$_3^+$ internal excitation.  The
range of possible ro-vibrational levels that can contribute to the
process effectively leads to a smearing out with relative
energy, unlike what was seen by \citet{Urba91a} for atomic collision
partners.  Additionally, the empirical fit to our data suggests that
the merged-beams rate coefficient peaks at around 1.37~eV.  This is
about 0.61~eV below the opening of the first competing pathway
at 1.98~eV, implying a level of internal excitation for the
  H$_3^+$ in our experiment that is in rough agreement with the
  predictions of \citet{Anic84a} and the measurements of Urbain
  (private communication), which are discussed in
    Section~\ref{sec:molecular}.  Using the partition function of
  \citet{Kyla11a}, this 0.61~eV of excitation corresponds to an
  internal temperature of $\sim 2,500$~K.

\subsubsection{$\mathrm{C+H_3^+ \rightarrow CH_2^+ + H}$}

Our results for reaction~(\ref{eq:CH3+_ppT}) show a decreasing
merged-beams rate coefficient with increasing relative energy.
\citet{Bett98a} describe the formation process of CH$_2^+$ as
involving the rearrangement of the CH$_3^+$ complex followed by the
ejection of one hydrogen atom.  Using this as the basis of a
  hand-waving argument, we attribute the observed energy dependence to
  the decreasing time available for the rearrangement of the CH$_3^+$
  complex as the collision energy increases.  Clearly, though, further
  theoretical and experimental work will be needed to resolve this
  issue.  Then at $\langle E_{\rm r}\rangle\sim 1$~eV, similarly to
that observed for reaction~(\ref{eq:CH3+_pT}), the process rapidly
decreases in strength. This suggests that we are seeing the onset of
the competitive channels, reactions~(\ref{eq:Diss1}) to
(\ref{eq:Diss3}), but again shifted to a lower relative energy
due to the internal excitation the H$_3^+$ in our measurement.

\subsection{Translational Temperature Rate Coefficients}
\label{sec.kineticrate}

The reaction of ground-term C with cations is predicted by
\citet{Gent77a} to be driven by the long-range shape of the PES.  At
these distances we do not expect the internal excitation of the
H$_3^+$ to play any role.  Based on these assumptions and the
statistical population of the C fine-structure levels, we expect our
translational temperature rate coefficient to be equivalent to a
thermal equilibrium rate coefficient for $T \gtrsim 300$~K.

\subsubsection{$\mathrm{C+H_3^+ \rightarrow CH^+ + H_2}$}

At 300~K, the Langevin rate coefficient adopted for this
reaction by the astrochemical databases \citep{KIDA12,UMIST13} is a
factor of $\approx 2.9$ times larger than our results, while the
semi-classical results of \citet{Talb91a} and \citet{Bett98a,Bett01a}
are larger by factors of $\approx 3.3$ and $\approx 1.8$,
respectively, as can be seen in Figure~\ref{Fig:thermalrate}.  The
cause for the discrepancies is not immediately obvious.  It is
unlikely to be due to the differences in the population of the
fine-structure levels in the atomic C, which are expected to be
statistically populated in gas above 300~K as well as in our
experiment.  So it seems to us that a more likely explanation for the
discrepancies is that the actual potential energy surfaces are less
attractive than those used in the calculations, possibly due to an
underestimate of the spin-orbit coupling strength \citep[see
  Equations~6-10, Figure~4, and Appendix~B of][]{Klip10a}.

One might also be tempted to attribute these differences to the
internal excitation of the H$_3^+$ ions used for the present results,
as the calculations were performed for internally cold H$_3^+$.
However, a comparison to the work of \citet{Savi05a} strongly suggests
that this is not the case.  Our translational temperature rate
coefficient is in very good agreement with the mass-scaled results of
\citet{Savi05a}. Their work used an effusive C beam, at an estimated 
translational temperature of $T_{\rm C} \sim 3,000$~K, colliding 
with D$_3^+$ stored in an ion trap, with wall temperatures of $T_{\rm
  D_3^+} = 77$~K. Assuming that the D$_3^+$ cloud is approximately at
rest with respect to the C beam, the translational temperature
of the interaction $T_{\rm t}$ is given by
\begin{equation}
\frac{3 k_\mathrm{B} T_{\rm C}}{2m_{\rm C}} =  
\frac{3 k_\mathrm{B} T_{\rm t}}{2\mu},
\end{equation}
yielding $T_{\rm t} \sim 1,000$~K. The very good agreement between
their work and ours suggests that at this translational
temperature the internal excitation of the H$_3^+$ affects our results
at a level constrained by the size of the mutual experimental
uncertainties.  We expect this to remain valid down to 300~K, as we
can posit no reason for this situation to change so long as the
fine-structure levels of the C remain statistically populated.  Thus,
at 300~K it seems unlikely that the differences between theory and our
results can be attributed to internal excitation of the H$_3^+$.

\subsubsection{$\mathrm{C+H_3^+ \rightarrow CH_2^+ + H}$}

Our results indicate that there is no energy barrier for
reaction~(\ref{eq:CH3+_ppT}) with internally excited H$_3^+$.  This is
to be contrasted with the calculations of \citet{Talb91a} who predict
the existence of such a barrier.  Later calculations by
\citet{Bett98a,Bett01a} find no such barrier, but their predicted
thermal rate coefficient lies a factor of $\approx 26.7$ below our
results at 300~K (see Figure~\ref{Fig:thermalrate}).  Results similar
to ours were found by \citet{Savi05a} for internally cold D$_3^+$.
The very good agreement that we find between our work and their
mass-scaled results suggests that indeed there is no barrier for this
particular system.

Coming back to the work of \citet{Savi05a}, they could not exclude
the possibility of the deuterium-abstraction parasitic reaction
\begin{equation}
\mathrm{CD}^+ + \mathrm{D_2} \rightarrow 
\mathrm{CD}_2^+ + \mathrm{D} 
\label{eq:CD2_para}\\
\end{equation}
contributing to the formation of CD$_2^+$. For that reason they gave
only lower limits for their uncertainty on
reaction~(\ref{eq:CD3+_pnpnT}). Such a parasitic reaction would
effectively reduce their inferred CD$^+$ rate coefficient while
boosting their CD$_2^+$ rate coefficient. In our experimental setup,
the low density of the parent H$_3^+$ beam combined with the low
CH$^+$ formation rate yields an insignificant rate for a parasitic
reaction forming CH$_2^+$. Hence, the good agreement between our
results for reactions~(\ref{eq:CH3+_pT}) and (\ref{eq:CH3+_ppT}) and
their mass-scaled results for reactions~(\ref{eq:CD3+_pnT}) and
(\ref{eq:CD3+_pnpnT}) suggest that, to within our mutual error bars,
parasitic reactions were not an issue for their measurements of these
two reactions.

\subsubsection{$\mathrm{C+H_3^+ \rightarrow CH_3^+ + \mathrm{photon}}$}

For radiative association to occur, the CH$_3^+$ collision complex
must radiate away binding energy plus any internal energy
of the parent H$_3^+$ ion.  Based on our results for
reaction~(\ref{eq:CH3+_RA}), we can put a $1\sigma$ upper limit on the
translational temperature rate coefficient at 72~K of $5.76
\times 10^{-11}$~cm$^3$~s$^{-1}$, for an H$_3^+$ internal energy of
$\sim 0.6$~eV.  The mass-scaled results of \citet{Savi05a} of $(5 \pm
3) \times 10^{-11}$~cm$^3$~s$^{-1}$, lie within this limit, though
their results are for the much higher translational temperature
of $\sim 1,000$~K and with insignificant internal excitation of the
H$_3^+$.  Still we find it unlikely that the rate coefficient for this
reaction can be as high as their results suggest.  Their data imply a
surprisingly flat temperature dependence for this radiative
association reaction.  Moreover, previous experimental and theoretical
studies for radiative association reactions have found rate
coefficients many orders of magnitude smaller \citep{Gerl92a}.
\citet{Savi05a} suggest that their results for this reaction may have
been contaminated by parasitic reactions. For example, there is the
two-step process of reaction~(\ref{eq:CD3+_pnpnT}) followed by the
deuterium-abstraction reaction
\begin{equation}
\mathrm{CD}_2^+ + \mathrm{D}_2 \rightarrow 
\mathrm{CD}_3^+ + \mathrm{D}, 
\label{eq:CD3_para}
\end{equation}
or the three-step, and therefore less likely, process of
reaction~(\ref{eq:CD3+_pnT}) followed by reaction~(\ref{eq:CD2_para})
and then by reaction~(\ref{eq:CD3_para}).  The rate coefficient
measured by \citet{Savi05a} can readily be explained if in their
apparatus the effective rate coefficient for either of these two
pathways was $\sim 10\%$ of the rate coefficient for the initial
step. Unfortunately the uncertainty limits on their and our results do
not enable us to tease out the explanation for their having measured
such a high rate coefficient for this radiative association process.

\subsection{Converting Translational Temperature to Thermal  
Rate Coefficients} \label{sec:kinetic2thermal}

\citet{Talb91a} and \citet{Bett98a,Bett01a} carried out their
calculations at 10~K.  In their work, they treat the reaction
  adiabiatically and ignore surface crossing and intersystem
  transitions.  This approach is still standard for theoretical rate
  coefficients \citep{Klip10a,Li14a}.  Additionally, 
  \citet{Talb91a} and \citet{Bett98a,Bett01a} extrapolated to
temperatures up to 300~K by multiplicatively scaling their results to
account for the temperature dependence of fractional populations of
the attractive surfaces involved in the reaction.  Here we take a
similar approach.  Using our translational temperature results
on statistically populated ground-term C, we convert them to thermal
rate coefficients using factors, derived below, which account for the
temperature dependence of the fractional population of the attractive
surfaces involved.  In converting our translational temperature
  results to thermal data, we also ignore the internal excitation of
  the H$_3^+$ in our experiment.  The role of internal excitation
  remains an open question for barrierless complex-forming reactions
  such as those studied here.  Typically no enhancement in reactivity
  is expected \citep[e.g.,][]{Guo12a}.  However, experimental and
  theoretical work has shown that for some reactions internal
  excitation can significantly enhance reactivity \citep[see][and
    references therein]{Li14a}.  Our work cannot resolve this issue.
  But ignoring the H$_3^+$ internal excitation here seems a reasonable
  approximation based on the good agreement between our results on hot
  H$_3^+$ and the mass-scaled results of \citet{Savi05a} on cold
  D$_3^+$.

\citet{Gent77a} investigated reactions of ground-term atomic C with
cations.  That work used the adiabatic approximation in which the
process is driven by the long-range shape of the PESs.  The only
aspect of the cation accounted for is the charge. They find that for
ground-term C reacting with a cation, the nine states in the
C($^3P_J$) manifold form six attractive surfaces and three repulsive.
These can be characterized at long range by the $J$ level and $|M_J|$
state of the carbon.  The one state of the $^3P_0$ level and three
states of the $^3P_1$ level correlate with attractive surfaces.  For
the $^3P_2$ level, two of the states correlate to attractive surfaces
and three to repulsive.

Building on the work of \citet{Gent77a}, \citet{Talb91a} and
  \citet{Bett98a,Bett01a} have extended it to reactions with $\rm
  H_3^+$.  They find that the reaction proceeds not through the
  singlet ground symmetry of CH$_3^+$, but rather through excited
  triplet surfaces of the intermediate ${\rm CH^+\cdot H_2}$.  In
  fact, in the reaction the C atom is predicted to be preferentially
  directed towards the apex of the H$_3^+$ triangle \citep[cf., Figure
    1 of][]{Bett98a} The lowest energy triplet is the
  $^3$A$^{\prime\prime}$, which does not lie along the ${\rm CH^+\cdot
    H_2}$ reaction path.  Complete rearrangement is needed to reach
  the $^3$A$^{\prime\prime}$ state, namely the insertion of the C atom
  between the three H nuclei.

Using the adiabatic approximation \citet{Talb91a} and
\citet{Bett98a,Bett01a} correlate the carbon fine-structure levels to
the PESs of the ${\rm CH^+\cdot H_2}$ intermediate, namely the
$^3$B$_2$ symmetry and the slightly higher $^3$B$_1$ symmetry, which
are both attractive, while the next higher triplet symmetry, the
$^3$A$_2$ is repulsive.  The six attractive $\mathrm{C+H_3^+}$
surfaces are assumed to correlate adiabatically with the six
attractive surfaces formed by the $^3$B$_2$ and $^3$B$_1$ symmetries
and the three repulsive $\mathrm{C+H_3^+}$ surfaces to the three
formed by the $^3$A$_2$ symmetry.

This leads to a one-to-one mapping of the long-range surfaces to those
of the CH$_3^+$ intermediate.  Using a $|J,|M_J|\rangle$ ket notation
for the ground-term C, the $|0,0\rangle$ state, the $|1,0\rangle$
state, and one of the two $|1,1\rangle$ states all map to the
$^3$B$_2$ symmetry.  The other one of the $|1,1\rangle$ states and
both of the $|2,2\rangle$ states all map to the $^3$B$_1$ symmetry.
Lastly, both of the $|2,1\rangle$ states and the single $|2,0\rangle$
state all map to the $^3$A$_2$ symmetry.

\begin{figure}[!p]
\centering
\includegraphics[width=1\textwidth]{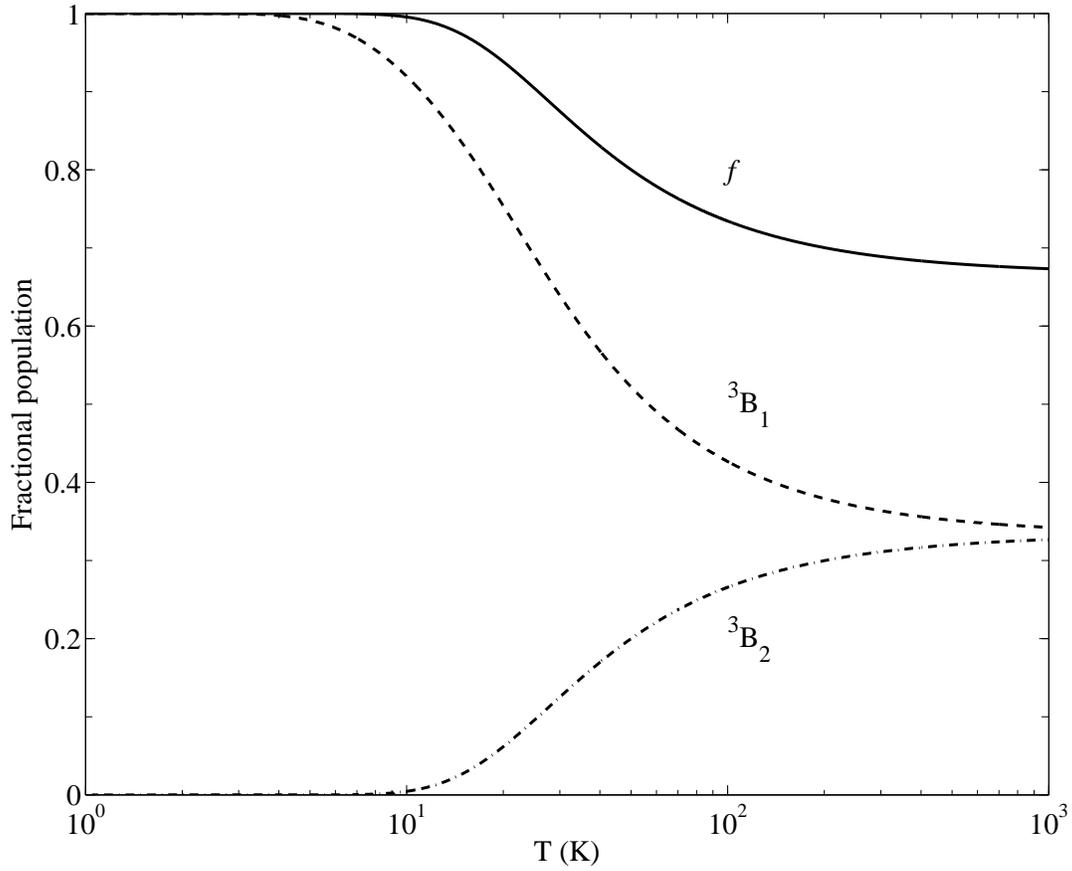} 
\caption{Fractional population of the two lowest CH$_3^+$ reactive triplet symmetries and their sum $f$ versus temperature.}
\label{fig:CH3+levels}
\end{figure}

Putting this all together, the partition functions for the attractive
$^3$B$_2$, $^3$B$_1$, and $^3$A$_2$ symmetries are given by
\begin{eqnarray}
u_{\rm ^3B_2} &=& u_0 + \frac{2}{3}u_1, \\
u_{\rm ^3B_1} &=& \frac{1}{3}u_1 + \frac{2}{5}u_2, \\
u_{\rm ^3A_2} &=& \frac{3}{5}u_2,
\end{eqnarray}
with $u_J$ defined by Equation~(\ref{eq:partition}).  The partition
functions for the $^3$B$_2$ and $^3$B$_1$ symmetries are shown in
Figure~\ref{fig:CH3+levels}.  
The temperature dependence for the fractional population on
attractive surfaces forming either of these ions is given by the
factor
\begin{equation}
f = u_{\rm ^3B_2} + u_{\rm ^3B_1},\label{eq:TRatios}
\end{equation}
which starts out at 1 at low temperature and decreases to 2/3 at high
temperature, as can be seen in Figure~\ref{fig:CH3+levels}.  The
partition function of the repulsive $^3$A$_2$ symmetry is given by
$1-f$.  All three partition functions converge to a value of $1/3$ at
high temperature, i.e., for statistically populated $J$ levels.  Based
on the above discussion and using these partition functions, we can
develop all the scale factors needed to account for the temperature
dependence of the fractional population on attractive surfaces.

\citet{Talb91a} and \citet{Bett98a,Bett01a} find that the $^3$B$_{2}$
and $^3$B$_{1}$ symmetries both lead to $\rm CH^+(a^3\Pi) + H_2$
formation.  Calculations by \citet{Bett98a,Bett01a} indicate that the
cross sections are the same for formation of CH$^+$ via the $^3$B$_2$
and $^3$B$_1$ symmetries.  Furthermore, they find that ground-symmetry
CH$_2^+(^2{\rm A}_1)$ forms only via the CH$_3^+$($^3$B$_2$) surface.

\begin{figure}[!p]
\centering
\includegraphics[width=1\textwidth]{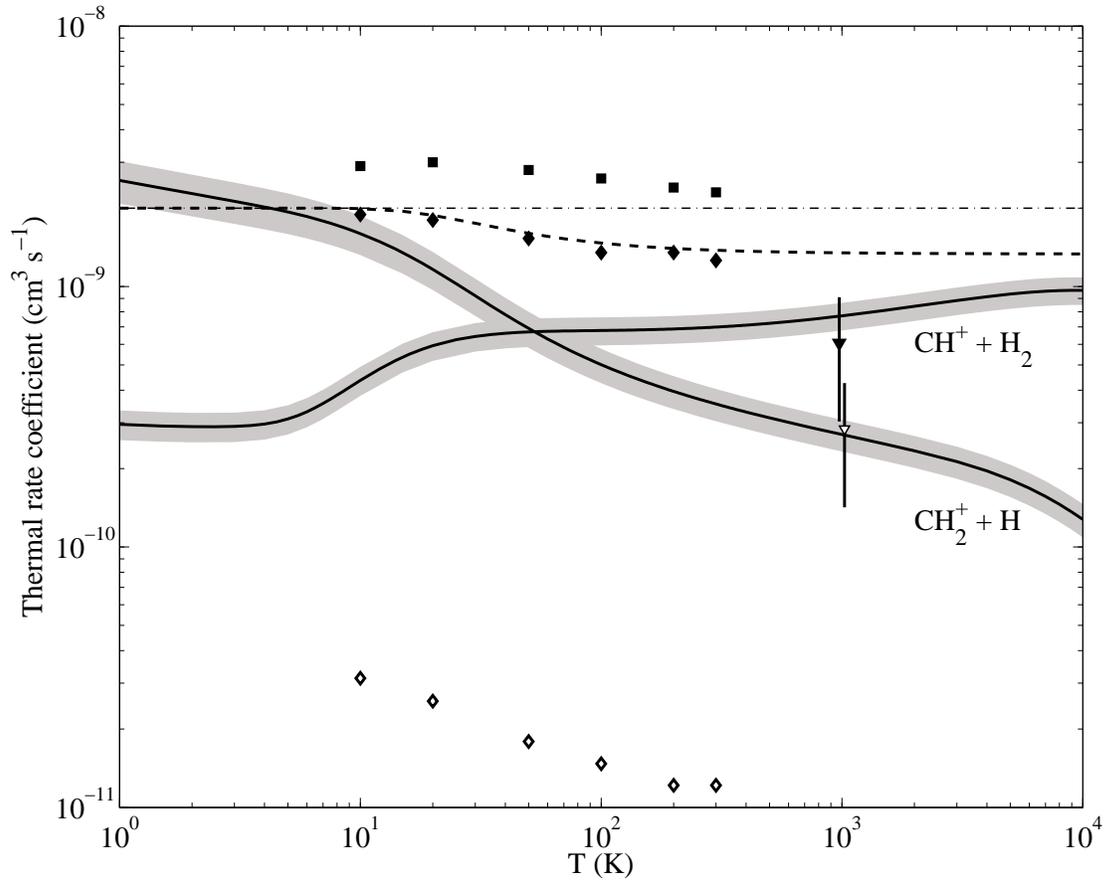}
\caption{The solid curves represents our experimentally derived
  thermal rate coefficients as described in
  Section~\ref{sec:kinetic2thermal}.  The shaded areas show the
  estimated $1\sigma$ total experimental uncertainty.  The dashed
  curve is the modified Langevin value, given by
  Equation~(\ref{eq:TLang}).  All other theoretical and experimental
  results are the same as in Figure~\ref{Fig:thermalrate}.}
\label{fig:correctedrates}
\end{figure}

The calculations of \citet{Bett98a,Bett01a}, the measurements of
\citet{Savi05a}, and our results, all find that both the CH$^+$ and
CH$_2^+$ channels are open.  Thus the Langevin rate coefficient
corresponds to the sum of the rate coefficients for these two
channels.  The temperature dependent Langevin thermal rate coefficient
can then be written as
\begin{equation}
\alpha_{\rm L}(T) = f(T)\alpha_{\rm L}(T=0), \label{eq:TLang}
\end{equation}
The resulting temperature dependent Langevin rate coefficient is shown
in Figure~\ref{fig:correctedrates}.

\begin{figure}[!p]
\centering
\includegraphics[width=1\textwidth]{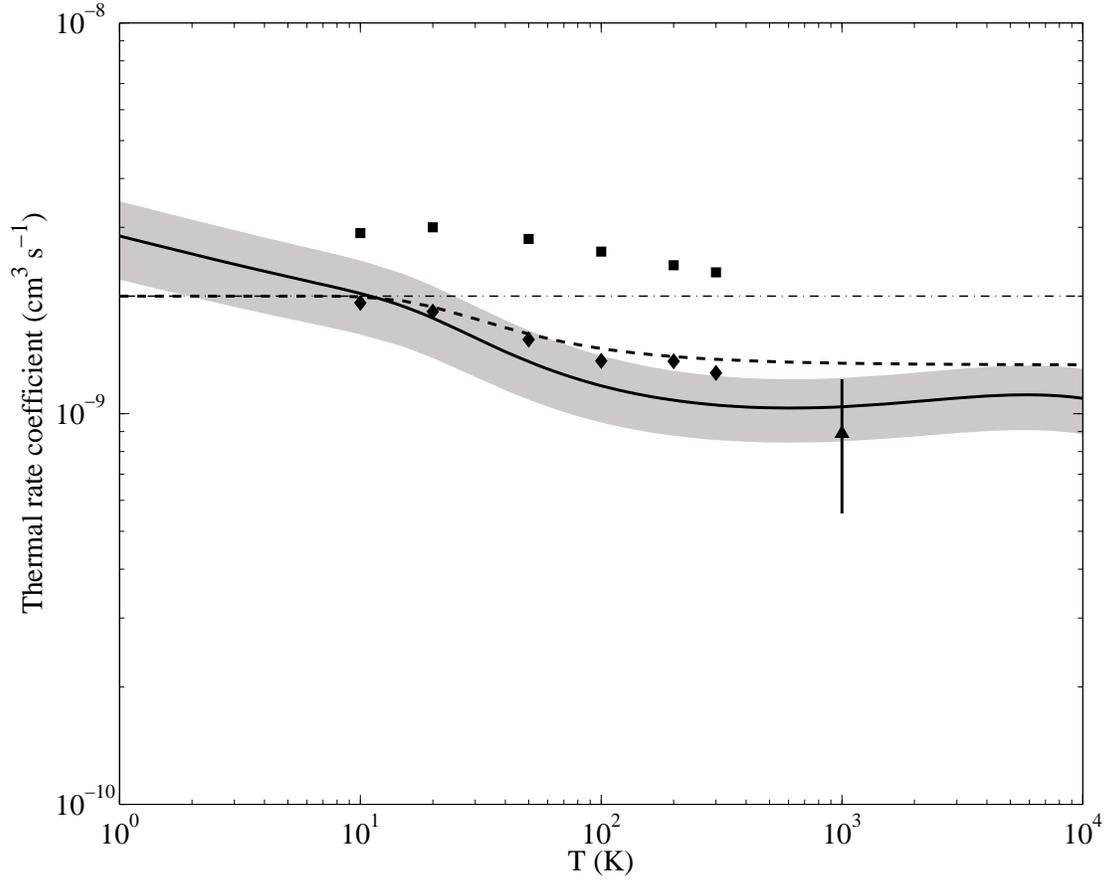}
\caption{Summed thermal rate coefficients for ${\rm C + H_3^+}$
  forming both CH$^+$ and CH$_2^+$.  The solid curve presents our
  experimentally derived results and the shaded area the quadrature
  sum of the errors shown in Figure~\ref{fig:correctedrates}.  The
  dot-dashed curve shows the unmodified Langevin rate coefficient and
  the dashed curve the modified value.  The squares present the
  theoretical results of \protect\citet{Talb91a} and the diamonds
  those of \protect\citet{Bett98a,Bett01a}.  The triangle gives the
  mass-scaled experimental results of \protect\citet{Savi05a}.}
\label{fig:correctedsummedrates}
\end{figure}

In order to convert our translational temperature results to
thermal rate coefficients, we first add together the   translational temperature rate coefficients for
reactions~(\ref{eq:CH3+_pT}) and (\ref{eq:CH3+_ppT}) and multiply the
sum by $\frac{3}{2}f$.  This corrects our experimental data where only
two-thirds of the C fine-structure levels contribute to the reaction
process.  Next we convert our results for CH$_2^+$ formation based on
the predictions of \citet{Bett98a,Bett01a} that only the $\rm ^3B_2$
symmetry is involved.  Hence, to derive the thermal rate coefficient
for reaction~(\ref{eq:CH3+_ppT}) we need only multiply the 
  translational temperature results by the factor $3 u_{\rm{^3B_2}}$.
This corrects our experimental data where only one-third of the C
fine-structure levels contribute to the formation of CH$_2^+$.
Lastly, for the thermal rate coefficient for
reaction~(\ref{eq:CH3+_pT}), we take the summed thermal rate
coefficient for reactions~(\ref{eq:CH3+_pT}) and (\ref{eq:CH3+_ppT})
subtract from it that for reaction~(\ref{eq:CH3+_ppT}).  The resulting
thermal rate coefficients for reactions~(\ref{eq:CH3+_pT}) and
(\ref{eq:CH3+_ppT}) are shown in Figure~\ref{fig:correctedrates} and
for the summed thermal rate coefficient in
Figure~\ref{fig:correctedsummedrates}.

\subsection{Thermal Rate Coefficients}

Our experimentally derived thermal rate coefficient for
reaction~(\ref{eq:CH3+_pT}) decreases with decreasing temperature.  A
comparison with the theoretical calculations of \citet{Talb91a} and
\citet{Bett98a,Bett01a}, shown in Figure~\ref{fig:correctedrates},
finds poor agreement in both the magnitude and temperature dependence.
At 10~K the calculations of \citet{Talb91a} are a factor of $\approx
6.6$ greater than the experimental results.  This discrepancy
decreases with increasing temperature and is $\approx 3.3$ at 300~K.
Over this same temperature range, the calculations of
\citet{Bett98a,Bett01a} lie a factor of $\approx 4.3$ above ours at
10~K and $\approx 1.8$ at 300~K.  The current astrochemical databases
\citep{KIDA12,UMIST13} use the Langevin rate coefficient for
reaction~(\ref{eq:CH3+_pT}).  The unmodified and modified Langevin
rate coefficients also do a poor job of reproducing our experimental
results for this channel.  This is not surprising since, as discussed
above, the Langevin rate coefficient  should be taken as
  representing the sum of the rate coefficients for
reactions~(\ref{eq:CH3+_pT}) and (\ref{eq:CH3+_ppT}).  Lastly, we note
that the modified Langevin rate coefficient closely matches the
calculations of \citet{Bett98a,Bett01a}.

For reaction~(\ref{eq:CH3+_ppT}), our experimentally derived thermal
rate coefficient increases with decreasing temperature.  The
theoretical calculations of \citet{Bett98a,Bett01a} show a roughly
similar temperature dependence, but differ significantly in the
magnitude of the rate coefficient.  At 10 and 300~K, their
calculations lie a respective factor of $\approx 51$ and 29 times below
our results.  Theory appears to greatly underestimates the
importance of this channel.

Above $\sim 300$~K the statistical fractional population in our
experiment of the ground term C($^3P_J$) closely matches a thermal
distribution.  A similar situation is expected for the $\sim 3,000$~K
carbon atoms in the \citet{Savi05a} measurements.  Hence, we expect in
both experiments that the corresponding CH$_3^+$ symmetries will be
thermally populated.  Thus, above 300~K our experimental rate
coefficient remains essentially unchanged from that presented in
Figure~\ref{Fig:thermalrate}.  As such, for formation of both CH$^+$
and CH$_2^+$, there remains good agreement of our results with the
mass-scaled experimental data from \citet{Savi05a}.

We attribute all the differences between theory and our experimental
results to a combination of factors.  The first, as mentioned
earlier, is that the potential energy surface of the ${\rm C + H_3^+}$
system is less attractive at large internuclear distances than
currently predicted.  Secondly, formation of CH$_2^+$ is over an order
of magnitude easier than calculated by published theory.  Lastly,
these effects are amplified by the temperature dependence for the
fractional population of reacting states.

\begin{figure}[!p]
\centering
\includegraphics[width=1\textwidth]{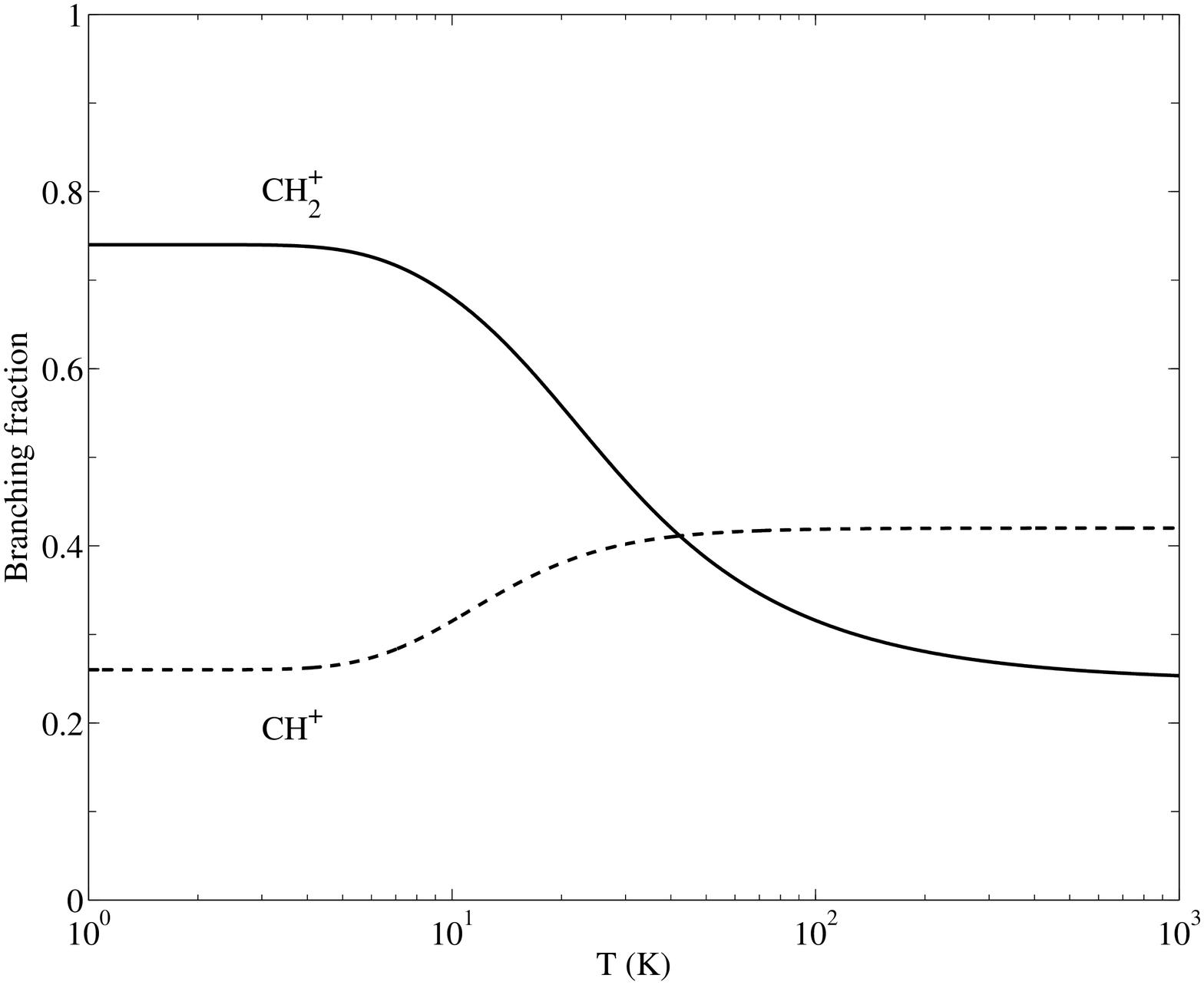} 
\caption{Temperature dependent branching fractions for the ${\rm C
      + H_3^+}$ reaction forming CH$^+$ and CH$_2^+$.}
\label{fig:branchingratios}
\end{figure}

The effects of the changing population of the attractive $^3$B$_1$ and
$^3$B$_2$ symmetries can also be seen in the
Figure~\ref{fig:branchingratios} which shows the branching ratio for
${\rm C + H_3^+}$ forming CH$^+$ and CH$_2^+$.  Formation of CH$^+$
dominates above $\sim 50$~K.  Below this temperature the populations
of the $^3$B$_1$ and $^3$A$_2$ symmetries decrease and all of the
population shifts into the $^3$B$_2$ symmetry, reducing the rate
coefficient for CH$^+$ formation and increasing that for CH$_2^+$
formation.  In fact, our results predict that at the $\sim 10$~K
temperature typical of dark molecular clouds approximately 80\% of all
${\rm C + H_3^+}$ reactions lead directly to CH$_2^+$.  This may be an
issue for astrochemical models as this important channel is currently
absent from astrophysical databases.

\section{Some Astrophysical Implications}
\label{sec:astro}

In gas-phase astrochemistry of dark molecular clouds, the CH$_3^+$
molecule is predicted to play a key role in the synthesis of complex
organic molecules \citep{Smit95}.  Reaction~(\ref{eq:CH3+_pT})
contributes to CH$_3^+$ formation via the hydrogen abstraction chain
\begin{equation}
{\rm CH^+ \stackrel{H_2}{\longrightarrow} CH_2^+ 
\stackrel{H_2}{\longrightarrow} CH_3^+} 
\label{eq:CH+_Hab}.
\end{equation}
On the scale of a dark cloud lifetime, the initial formation of CH$^+$
via reaction (\ref{eq:CH3+_pT}) is slow due to the low abundances of C
and H$_3^+$ in the cloud.  However, once the CH$^+$ molecule is formed
it rapidly proceeds to CH$_3^+$ due to the high H$_2$ abundance and
fast rate coefficient for the hydrogen abstraction reactions.  Our
thermal rate coefficient for reaction~(\ref{eq:CH3+_pT}) at 10~K is
significantly smaller than the unmodified Langevin value used in
current astrochemical databases \citep{KIDA12,UMIST13}.  At higher
temperatures, it is still significantly reduced, suggesting a slower
formation rate at higher temperatures for CH$_3^+$ from the ${\rm C +
  H_3^+}$ pathway and thereby a reduced abundance of more complex
organic molecules.

One needs, however, to also take into account
reaction~(\ref{eq:CH3+_ppT}) which is currently not included in
astrochemical models. This can lead to the formation of CH$_3^+$ via
\begin{equation}
{\rm CH_2^+ \stackrel{H_2}{\longrightarrow} CH_3^+},
\label{eq:CH2+_Hab}
\end{equation}
which also proceeds rapidly for the same reasons as reaction
pathway~(\ref{eq:CH+_Hab}).  We note reaction~(\ref{eq:CH2+_Hab}) also
enables CH$_2^+$ formed via reaction~(\ref{eq:C+H2_RA}) to go on to
form CH$_3^+$. Thus the rate coefficient for the reaction complex
${\rm C + H_3^+}$ forming CH$_3^+$ is effectively the sum of the
thermal rate coefficients for reactions~(\ref{eq:CH3+_pT}) and
(\ref{eq:CH3+_ppT}).

Figure~\ref{fig:correctedrates} shows the summed thermal rate
coefficients for our results as well as the published theoretical and
experimental results.  Table~\ref{tab:thermalratecompare} provides a
numerical comparison at selected temperatures.  The results of
\citet{Talb91a} always fall outside the estimated $1\sigma$
experimental uncertainty limits.  The unmodified Langevin rate
coefficient lies within these experimental limits from $\sim 2$ to
30~K.  The modified Langevin results show reasonable agreement from
$\sim 2$ to 60~K and are only slightly discrepant above $\sim 60$~K.
Comparing to the summed results of \citet{Bett98a,Bett01a}, we find
surprisingly good agreement over most of the $10-300$~K temperature
range that they covered.  Lastly, we continue to find good agreement
with the experimental results of \citet{Savi05a}.

Based on these comparisons, we expect that incorporating our results
into astrochemical models will have a temperature-dependent
effect on the gas-phase formation rate for CH$_3^+$ and the more
complex organic molecules that are formed from CH$_3^+$.  At 10~K our
summed rate coefficient for the reaction is in good agreement with the
unmodified Langevin value currently used in the astrochemical
databases.  However, at 300~K, our summed value is a factor of $\sim
2$ lower.  Determining the full astrochemical implications of our
results will require a detailed chemical simulation incorporating our
findings, which is beyond the scope of our work here.

We also would like to point out that in molecular clouds, for
densities of $10^4$~cm$^{-3}$ or higher, the population of the $^3P_J$
levels is expected to be in thermal equilibrium.  At lower densities
the excitation of the $J$ levels is probably less than thermal
\citep{Nuss79a}.  It is clear that the field will eventually need
fine-structure resolved rate coefficients.

\section{Summary}
\label{sec:summary}

We have developed a novel, merged fast-beams apparatus which allows us
to merge a beam of molecular ions onto a neutral beam of ground-term
atoms.  Here we have described the apparatus in detail.  For the
proof-of-principle studies, we have measured the chemistry of ${\rm C
  + H_3^+}$ forming CH$^+$, CH$_2^+$, and CH$_3^+$.  Our measurements
were performed for statistically populated C($^3P_J$) in the ground
term, which is nearly equivalent to the population expected for
thermal temperatures above 300~K.  Hence, our translational
temperature results are expected to be similar to thermal results
above this temperature.  At $\sim 1,000$~K, we find good agreement
between our results and the mass-scaled results from published ion
trap measurements for ${\rm C + D_3^+}$ forming CD$^+$ and CD$_2^+$.
At 300~K, our results for CH$^+$ formation lie a factor of $\sim
1.8-3.3$ below both the unmodified Langevin value currently in the
astrochemical databases and the published semi-classical results.
These databases do not currently include the CH$_2^+$ formation
channel.  Our translational temperature results at 300~K for
forming CH$_2^+$ are a factor of $\approx 26.7$ larger than the
semi-classical results.  Additionally, we have used statistical
arguments as a guide to convert our translational temperature
results to thermal results for temperatures below 300~K.  Our
conversion indicates that formation of CH$_2^+$ will dominate over
that of CH$^+$ at temperatures below $\sim 50$~K.  Clearly, though,
further experimental work using cold H$_3^+$ molecules and more
sophisticated theoretical calculations are needed to test this
prediction.

\section{Acknowledgments}

The authors thank E.\ Herbst, M.\ Delsaut, J.\ Li\'evin,
B.\ J.\ McCall, E.\ F.\ McCormack, and P.\ C.\ Stancil for stimulating
conversations.  We also thank the referee for their insightful
  and helpful review.  This work was supported in part by the
Advanced Technologies and Instrumentation Program and the Astronomy
and Astrophysics Grants Program in the NSF Division of Astronomical
Sciences.  X.\ Urbain is Senior Research Associate of the F.R.S.-FNRS.

\bibliography{CH3plus}
\clearpage

\begin{sidewaystable}[!pc]
\centering
\caption{Summary of typical statistical-like uncertainties going into
  Equation~(\ref{eq:rate}) for a single data run $i$.  Also listed are
  the relevant symbols and their typical values.  All uncertainties
  are quoted at a confidence level taken to be equivalent to a
  1$\sigma$ statistical confidence level, treated as random sign
  errors, and added in quadrature. \label{Tab:Values1}}
\begin{tabular}{lcccc}
\hline
\hline
Source & Symbol & Section & Value & Uncertainty (\%) \\
\hline 
Signal rate     & $S$                      	 & \ref{sec:Signal}                  & 1-15 Hz                           & $\leq$ 4\\
C velocity      & $v_\mathrm{n}$             & \ref{sec:carbon}                  & $6.7 \times 10^7$~cm~s$^{-1}$ & $\ll$1   \\
H$_3^+$ velocity & $v_\mathrm{i}$	  				 & \ref{sec:molecular}               & $6.7 \times 10^7$~cm~s$^{-1}$ & $\ll$1   \\
C current       & $I_\mathrm{n}$             & \ref{sec:neutralcurrent}          & 30 nA                             & 5        \\
H$_3^+$ current	& $I_\mathrm{i}$	           & \ref{sec:interaction}             & 200 nA                            & 5        \\
Overlap factor  & $\left\langle\Omega(z)\right\rangle$ 	 &\ref{sec:energies}                 &2.7 cm$^{-2}$                      & 10       \\
\hline
Statistical-like uncertainty (single run) & & & &  13 \\
\hline
\hline
\end{tabular}
\end{sidewaystable}

\clearpage
\begin{table}[!p]
\centering
\caption{Same as Table \ref{Tab:Values1} but for the systematic uncertainties for all runs.\label{Tab:Values2}}
\begin{tabular}{lcccc}
\hline 
\hline
Source & Symbol & Section & Value & Uncertainty (\%) \\
\hline
Analyzer transmission                   & $T_{\rm a}$ 		& \ref{sec:detector}       & 0.73     & 3 \\
Grid transmission                       & $T_{\rm g}$ 		& \ref{sec:detector}       & 0.90     & 1 \\
Neutral transmission                    & $T_{\rm n}$ 		& \ref{sec:neutralcurrent} & 0.94     & 2 \\
Neutral detector calibration 						& $\gamma$ 				& \ref{sec:neutralcurrent} & 2.7     & 11 \\
CEM efficiency       										& $\eta$          & \ref{sec:cemefficiency}  & 0.99     & 3 \\
Interaction length    									& $L$             & \ref{sec:interaction}    & 121.5 cm & 2 \\
\hline 
Total systematic uncertainty & & & & 12\\
\hline
\hline
\end{tabular}
\end{table}

\clearpage
\begin{table}[!p]
\centering
\caption{Fit parameters for the cross sections of reactions~(\ref{eq:CH3+_pT}) and (\ref{eq:CH3+_ppT}) vs.\ relative energy. The resulting cross sections using Equation~(\ref{Eqn.rateCH}) are in units of cm$^2$ for $E$ given in eV.}
\label{tab:rate_CH}
\begin{tabular}{cccccc}
\hline \hline
Reaction &\multicolumn{5}{c}{Parameters}\\
\cline{2-6}
			& $a_0$		& $a_{1/2}$ 	& $b_{1}$ 	& $b_{2}$ 	& $b_{4}$ \\
\hline
(\ref{eq:CH3+_pT}) 	&2.3474E-16 	& 1.1028E-15 	&	-	& 1.4694E-01    & 2.0471E-03\\
(\ref{eq:CH3+_ppT}) 	&1.9983E-16 	&	-	& 5.4737E-02    & 5.6944E-03    & 2.2891E-01\\
\hline \hline
\end{tabular}
\end{table}

\clearpage
\begin{table}[!p]
\centering
\caption{Fit parameters for the translational temperature rate
  coefficients for reactions~(\ref{eq:CH3+_pT}) and
  (\ref{eq:CH3+_ppT}). The resulting rate coefficients from
  Equation~(\ref{Eqn.thermalrateCH}) are in units of cm$^3$~s$^{-1}$
  for $T$ given in K.}
\label{tab:thermal_CH}
\begin{tabular}{ccccccc}
\hline \hline
Reaction &\multicolumn{6}{c}{Parameters}\\
\cline{2-7}
														& $a_0$				& $a_{1/2}$ 	& $a_{1}$ 				& $b_{1/2}$ 		& $b_{1}$ 			& $b_{3/2}$ \\
\hline
(\ref{eq:CH3+_pT})  &1.0218E-09		&	7.2733E-11		&	5.9203E-14		& 4.4914E-02	& -2.6056E-04		& 2.6397E-06	\\
(\ref{eq:CH3+_ppT})  &8.5145E-10		&	-							&	-							& 9.5666E-04	& -4.4040E-05		& 2.3496E-06 \\
\hline \hline
\end{tabular}
\end{table}

\clearpage
\begin{sidewaystable} [!p]
\centering
\caption{Summed thermal rate coefficients for ${\rm C + H_3^+}$
  forming both CH$^+$ and CH$_2^+$ for selected temperatures.  The
  rate coefficients are given in units of $10^{-9}$~cm$^3$~s$^{-1}$.}
\label{tab:thermalratecompare}
\begin{tabular}{cccccccc}
\hline \hline
Source & \multicolumn{7}{c}{Temperature (K)} \\
\cline{2-8}
	& 10 & 20 & 50 & 100 & 200 & 300 & 1000 \\
\hline
Langevin (unmodified)   & 2.0 & 2.0 & 2.0 & 2.0 & 2.0 & 2.0 & 2.0 \\
Langevin (modified)     & 2.0 & 1.9 & 1.6 & 1.5 & 1.4 & 1.4 & 1.3 \\
\citet{Talb91a}         & 2.9 & 3.0 & 2.8 & 2.6 & 2.4 & 2.3 & \\
\citet{Bett98a,Bett01a} & 1.9 & 1.8 & 1.5 & 1.4 & 1.4 & 1.3 & \\
\citet{Savi05a}         &     &     &     &     &     &     & $0.9 \pm 0.3$\\
Present results         & $2.0 \pm 0.4$ & $1.8 \pm 0.4$ & $1.4 \pm 0.3$ & $1.2 \pm 0.2$ & $1.1 \pm 0.2$ & $1.1 \pm 0.2$ & $1.0 \pm 0.2$\\
\hline \hline
\end{tabular}
\end{sidewaystable}

\clearpage

\clearpage

\clearpage

\clearpage

\clearpage

\clearpage

\clearpage

\clearpage

\clearpage

\clearpage

\clearpage

\clearpage

\clearpage


\clearpage

\clearpage

\clearpage

\clearpage

\end{document}